\documentclass{sig-alternate} 
\usepackage{mathptmx} 

\newcommand{\ignore}[1]{}
\usepackage{fancyhdr}
\usepackage[normalem]{ulem}
\usepackage[hyphens]{url}
\usepackage{hyperref}

\usepackage[normalem]{ulem}
\usepackage{morefloats}
\usepackage{graphics}
\usepackage{epsfig}
\usepackage{amsmath}
\usepackage{xspace}
\usepackage{enumitem}
\usepackage{enumerate}
\usepackage{mdwlist}
\usepackage{epstopdf}
\usepackage{textcomp}
\usepackage{xcolor}

\usepackage{fancyhdr}
\usepackage{listings}
\usepackage{listings}

\newcommand{\parbold}[1]{{\bf #1}}
\newcommand{\name}{E{\small MMA}\xspace}

\newcommand{\todo}[1]{}
\newcommand{\mohit}[1]{}
\newcommand{\mikhail}[1]{}
\newcommand{\ling}[1]{}

\begin{document}
\lstdefinestyle{customc}{
  belowcaptionskip=1\baselineskip,
  numbers=left,
  stepnumber=1,
  breaklines=true,
  frame=L,
  xleftmargin=\parindent,
  language=Java,
  showspaces=false,
  showstringspaces=false
  basicstyle=\ttfamily\tiny, 
  keywordstyle=\bfseries\color{green!40!black},
  commentstyle=\itshape\color{purple!40!black},
  identifierstyle=\color{blue},
  stringstyle=\color{orange},
}
\lstset{escapechar=@,style=customc}

\title{
\vspace{-0.5in} 
EMMA: A New Platform to \\
Evaluate Hardware-based Mobile Malware Analyses
}
\author{Mikhail Kazdagli ~~~~ Ling Huang$^{*}$ ~~~~ Vijay Reddi ~~~~ Mohit Tiwari \\[6pt]
University of Texas at Austin ~~~~~~~~~~~~~~~~~~~~ $^{*}$DataVisor Inc ~~~~~~ \\[6pt]
mikhail.kazdagli@utexas.edu ~~~~~~~~~~~~ hling@cs.berkeley.edu \\[3pt]
~~~~~~~~ vj@ece.utexas.edu ~~~~~~~~~~~~~~~~~~~~ tiwari@austin.utexas.edu
}

\date{}
\maketitle

\begin{abstract}
\sloppy

Hardware-based malware detectors (HMDs) are a key emerging technology
to build trustworthy computing platforms, especially mobile 
platforms. Quantifying the efficacy of HMDs against
malicious adversaries is thus an important problem.  The challenge lies in that
real-world malware typically adapts to defenses, evades being run in
experimental settings, and hides behind benign applications. Thus, realizing
the potential of HMDs as a line of defense -- that has a small and
battery-efficient code base -- requires a rigorous foundation for evaluating HMDs. 

To this end, we introduce \name---a platform to evaluate the efficacy of HMDs
for mobile platforms.  \name deconstructs malware into atomic, orthogonal
actions and introduces a systematic way of pitting different HMDs against a
diverse subset of malware hidden inside benign applications.  \name drives both
malware and benign programs with real user-inputs to yield an HMD's effective
\emph{operating range}---i.e., the malware actions a
particular HMD is capable of detecting. We show that small atomic actions, such
as stealing a Contact or SMS, have surprisingly large hardware footprints, and
use this insight to design HMD algorithms that are less intrusive than prior
work and yet perform 24.7\% better.  Finally, \name brings up a surprising new result---obfuscation
techniques used by malware to evade static analyses makes them more detectable using HMDs.

\ignore{



Evaluating hardware-based malware detectors (HMDs) is a hard problem---malware
adapts to defenses, evades being run in experimental settings, and hides behind
benign applications. Since adapting hardware-level computation is much simpler
than hiding system-level actions such as opening a socket or reading a file, 
realizing the potential of HMDs as a small and battery-efficient line of
defense 
requires a new, rigorous foundation for evaluating HMDs.

We introduce \name, a platform that deconstructs malware into atomic,
orthogonal actions and pits HMDs against adaptive malware hidden inside benign
applications. \name then drives both malware and benign programs with real
user-inputs to yield an HMD's effective operating range.  We find
that small atomic actions such as stealing a Contact or SMS have surprisingly
large hardware footprints, and use this data to create two new HMD algorithms
that are simpler and less intrusive than prior work but with comparable
performance.  Finally, we show that malware techniques to evade static analysis tools
make them surprisingly detectable using HMDs. 
\todo{update: case studies?}


%
\ignore{
payloads should be diversified to pin-point where malware detectors break down,
and that to determine its detection rate and false positives, malware should be compared to
its closest benign app while both apps driven by real user inputs.
In summary, this paper lays a solid methodological foundation for future research into 
novel hardware-based malware detectors.

Malware binaries are
hard to run correctly, can adapt their computational characteristics, and
require benign applications to be executed realistically for an appropriate
comparison.
We introduce \name, a novel methodology to evaluate mobile malware detectors that
aims to determine the {\em break-even point} of a detector; we designed a
malware benchmark suite to learn when malware becomes [un]detectable , as opposed
to treating malware as code-named black-boxes and reporting results solely as
true positive percentages. 
%
\name also includes a record-and-replay system to test detectors with real
user-driven executions of both benign and malware programs.

\ignore{

To demonstrate the utility of \name, we design and evaluate two hardware-based
anomaly detection algorithms, a probabilistic Markov model that accounts for
temporal behaviors and a bag-of-words model that doesn't.
Surprisingly, small malware payloads that steal a few SMSs can be detected as
computational anomalies even within complex apps such as Internet Radio,
AngryBirds etc. 
We also find that techniques malware use to obfuscate control and data flows
from static program analyses make their hardware signals {\em more} anomalous.
%
Finally, several hours of user-driven executions reveal that false positives
are a concern -- almost 7\% per 30s of benign execution -- and motivates
further research to turn
hardware detectors into more than just triggers for deeper emulator-based
forensics.
This work takes a step towards a sound methodological foundation for evaluating
novel hardware-based malware detectors.
}

}

}

\end{abstract}

\section{Introduction}
\label{sec:intro}
Hardware-based malware detectors (HMDs) are an attractive line of defense
against malware~\cite{simha-isca,anomaly-detector-columbia-raid,map,ponomarev-ensemble-learning}.  An HMD
extracts instruction and micro-architectural data from a program run 
and raises an alert when the current trace's statistics looks anomalous
compared to benign traces (or similar to a known malicious one). HMDs are small 
and can run securely even from a compromised OS---they are thus a trustworthy
first-level detector in a {\em collaborative} malware detection
system~\cite{cids-survey,cids-Zhou2010} and are being deployed in commercial mobile
devices.

Evaluating HMDs for mobile malware, however, is a new challenge for architects.
Unlike SPEC programs, malware only runs under specific conditions---on real
devices in select geographical regions triggered by commands from a remote
server.  
Without a malware benchmark suite, it is challenging to
experiment with a carefully diversified set of malware.
Further, HMDs have to differentiate malware from 
benign programs---without real inputs that cover 
a representative range of benign traces,
mobile
apps are quiet and HMDs will simply learn to label {\em any} computation as
malware. HMDs today are evaluated in a `black-box' manner --
without explicitly triggering malicious payloads and by comparing malicious
traces to quiescent benignware traces~\cite{simha-isca} --
such that neither malware nor benignware traces represent a
real execution. 

\ignore{
Furthermore, {\em desktop} malware is primarily exploit-driven, and we show
that techniques to evaluate desktop HMDs~\cite{simha-raid,dmitry-hpca} do not
port over to evaluating {\em mobile} malware.}


\begin{figure}[tbp]
   \centering
   \includegraphics[width=0.45\textwidth]{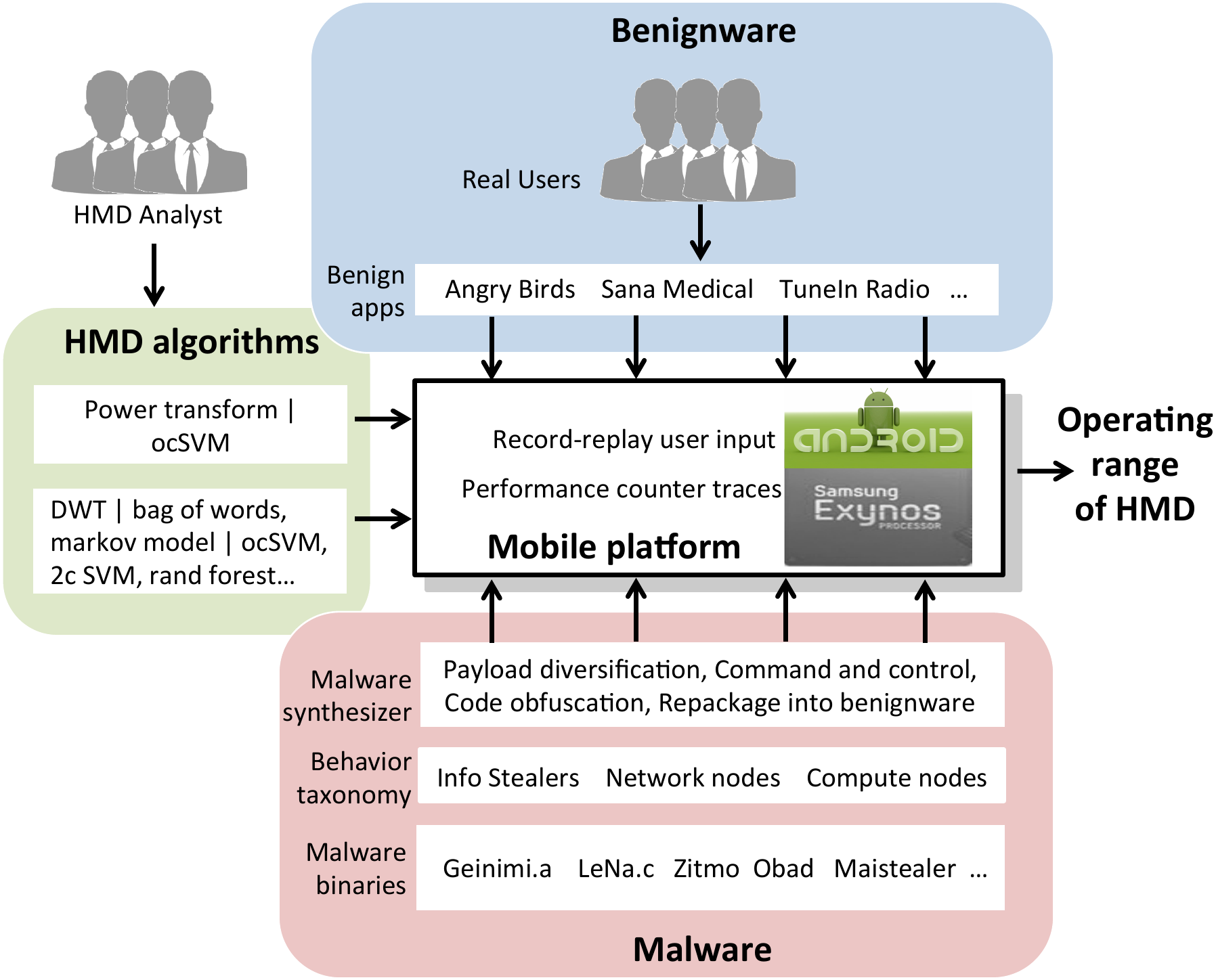}\par
   \caption{Overview of \name. 
}
\label{fig:sys-overview}
\end{figure}

In this paper, we present \name---a principled methodology to evaluate
HMDs for mobile malware (Figure~\ref{fig:sys-overview}). As a baseline advance over
prior work~\cite{simha-isca}, we reverse engineer real malware to execute
correctly and drive mobile apps using real human input on actual hardware that
contains realistic data.  We have built a custom record-and-replay framework
for Android apps to replay thousands of 5 to 10 minute long user interactions --
such as playing Angry Birds or filling out a medical diagnostic questionnaire
-- correctly.  Further, we explicitly model malware adapts its hardware level
behaviors to evade detection.  To this end, we present a taxonomy of real
malware into orthogonal behaviors (and atomic actions for each behavior) and
synthesize a diverse range of malware actions. 

\name helps a malware analyst find the {\bf operating range} of HMD algorithms.
An operating range is a new metric of the form: {\em an HMD algorithm A can
detect malware payload X hidden in app Y with a false positive rate of Z}. 
In contrast, HMDs' performance today is quantified using {\em
Receiver Operating Curves} (ROC plots) that show aggregate true positive v.
false positive rates across a suite of malware and benignware programs.
Aggregate ROCs are misleading because (a) adversaries can adapt payloads arbitrarily
{\em in response to the proposed HMD} -- hence, operating range is defined 
in terms of atomic malware payload units instead of true positive percentages
in ROC plots
-- and (b) false positives should be
measured using the benign app that malware hides in---comparing to an
arbitrary benign app or system utility yields an unrealistic (and better) false
positive rate.

We demonstrate \name's utility through three case studies that yield new
conclusions.  Our first case study shows that {\em anomaly-based} HMDs, that
flag novel executions as malware, benefit from \name's characterization of
atomic malware actions. Specifically, we find that desktop HMDs designed to
detect short-lived exploits are a poor fit to detect mobile malware payloads.
Further, small software level actions such as stealing a 4MB photo or one SMS
takes 2.86s and 0.12s respectively on a Samsung Exynos 5250 device. Using this
insight, we propose an HMD that uses longer-duration (100ms) feature vectors
and is 24.7\% more effective using the area under the ROC curve (AUC) metric
than prior work (at the same false positive rate of $\sim$20\%).

Our second case study uses \name's malware taxonomy to design effective {\em
supervised learning based} HMDs, i.e. HMDs trained on both benignware and known
malware.  We show quantitatively that supervised learning HMDs benefit from
training on a malware set that covers diverse, orthogonal behaviors (compared
to HMDs trained on a subset of behaviors).
Further, the supervised learning model can classify even small pieces of data
(1 photo, 25 contacts, 200 SMSs, etc) being stolen with close to 100\% accuracy
at 5\% false positive rate. However, malware payloads such as HTTP-layer denial
of service attacks are undetectable at the hardware level---\name provides such
semantic insights into why HMDs succeed and fail.

\ignore{For both detectors, example of a threshold in terms of SMSs, contacts, etc
to highlight operating range.}

Our final case study shows a surprising result---{\em obfuscation techniques to
evade static analysis tools make HMDs more effective.} Specifically, malware
developers use string encryption and Java reflection to create high-fanout
nodes in data- and control-flow graphs and thus foil static
analysis tools. However, these obfuscation techniques in turn create
instruction sequences and indirect jumps that make malware stand out from
benignware. 
Hence, in addition to collaborative malware detectors, light-weight HMDs can
complement static analysis tools~\cite{bouncer} used by Google and other app
stores to drive malware down into more inefficient design points.

\noindent To summarize, our specific contributions include:\\
\noindent{\bf 1. Malware taxonomy.}
We deconstruct 229 malware binaries from 126 families into orthogonal behaviors, 
identify atomic actions for each behavior, and build a malware 
synthesizer that incorporates state-of-the-art obfuscation and
command-and-control protocols.
We find that small software-level actions have large hardware footprints
and use this to design effective HMDs.

\noindent{\bf 2. Record and replay platform.}
We record real (human) user traces for 9 complex and popular
applications such as Angry Birds running on actual hardware with realistic data -- $\sim$1 to 2 hours for each app -- and show that these are very different from 
traces produced with none or auto-generated inputs.
We repackage the 9 benign apps into a total of 594 diverse malware binaries and 
replay over 4000 minutes of malware binaries to extract malicious payloads'
time intervals. We use this platform to evaluate HMD algorithms.




\noindent{\bf 3. Three case studies with new insights.}
Anomaly detectors, if tuned to atomic actions in real malware,
improve over prior HMDs by 24.7\%. 
Supervised-learning HMDs improve by 6--10\% if the training
set includes each high-level behavior from \name's taxonomy,
and can detect even small data items being stolen from within complex apps.
Finally, HMDs detect
what static analyses cannot---reflection and string encryption
improves our HMD's detection rate. 

\name has already informed the design and evaluation of a commercial
malware detector and is in use by an external academic 
research group. We will release the user traces, malware and 
benignware dataset, and the hardware platform
to researchers to seed composable research on HMDs.
Before we dive into the details of \name in Sections~\ref{sec:malware-gen}
and~\ref{sec:impl}, we motivate our approach by demonstrating how 
prior `black-box' approaches  to evaluating HMDs can lead to 
misleading results.

\ignore{
\noindent{\bf Insights}

- study payloads, not exploits
- payloads into orthogonal atomic actions
- suprisingly large footprint, that indicate potential
- Differential analysis: correct benign and malware
- Outcome: example 'Detector exposes payload-x hidden in app-y' instead of 'Detector catches 75\% of all malware'

\noindent{\bf Notes}

2nd para: collaborative IDS => trustworthy detector that is good enough to
raise a few TPs.

3rd: mimicry attacks

Questions for future work
- do Android programs have phases?
}



\section{Motivation}
\label{sec:motivation}
%


We consider HMDs as part of a collaborative malware detection system that has
two components.  On the server side, a platform provider (e.g., Google)
executes benign and/or malware applications using test and real user inputs,
measures performance counters, and creates a database of computational models.
On client devices, a light-weight {\em local detector} samples performance
counters to create run-time traces from applications, and compares each
run-time trace to database entries on the device and forwards suspicious traces
to a {\em global detector} on the server.

HMDs can build databases of {\em signatures} of both malware and benign
executions~\cite{simha-isca} or train only on benign executions to
flag {\em anomalous} executions as
malware~\cite{anomaly-detector-columbia-raid}---\name can be used to evaluate
both these classes of HMDs.
In a signature-based analysis, the HMD has to compare each run-time trace with
the entire database looking for a possible match. 
In an anomaly detector, each run-time trace purports to belong to a specific
app -- hence the HMD needs to match the current trace to only that specific app's 
model.
If malware is detected with high confidence, the global detector 
raises an alert to the user and/or a malware analyst. 

\begin{figure}[tbp]
   \centering
   \includegraphics[width=0.45\textwidth]{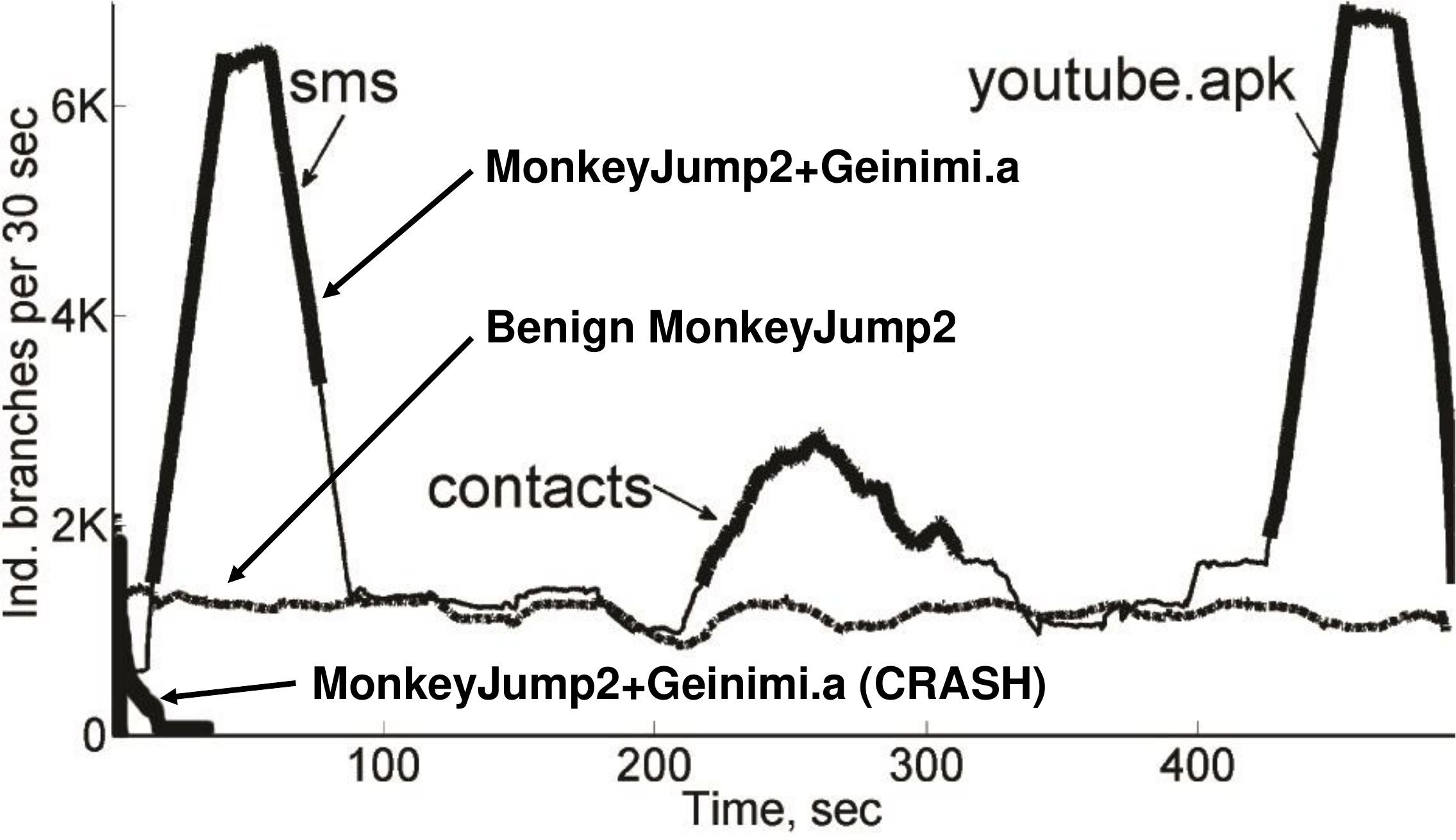}\par
   \caption{Executing malware payloads. The off-the-shelf Geinimi.a malware 
    crashes immediately. 
    Once fixed, Geinimi.a executes malicious payloads such as stealing
    SMSs or contacts or downloading files.	
    }
\label{fig:geinimi-crash}
\end{figure}

\begin{figure}
\centering
\includegraphics[width=0.45\textwidth]{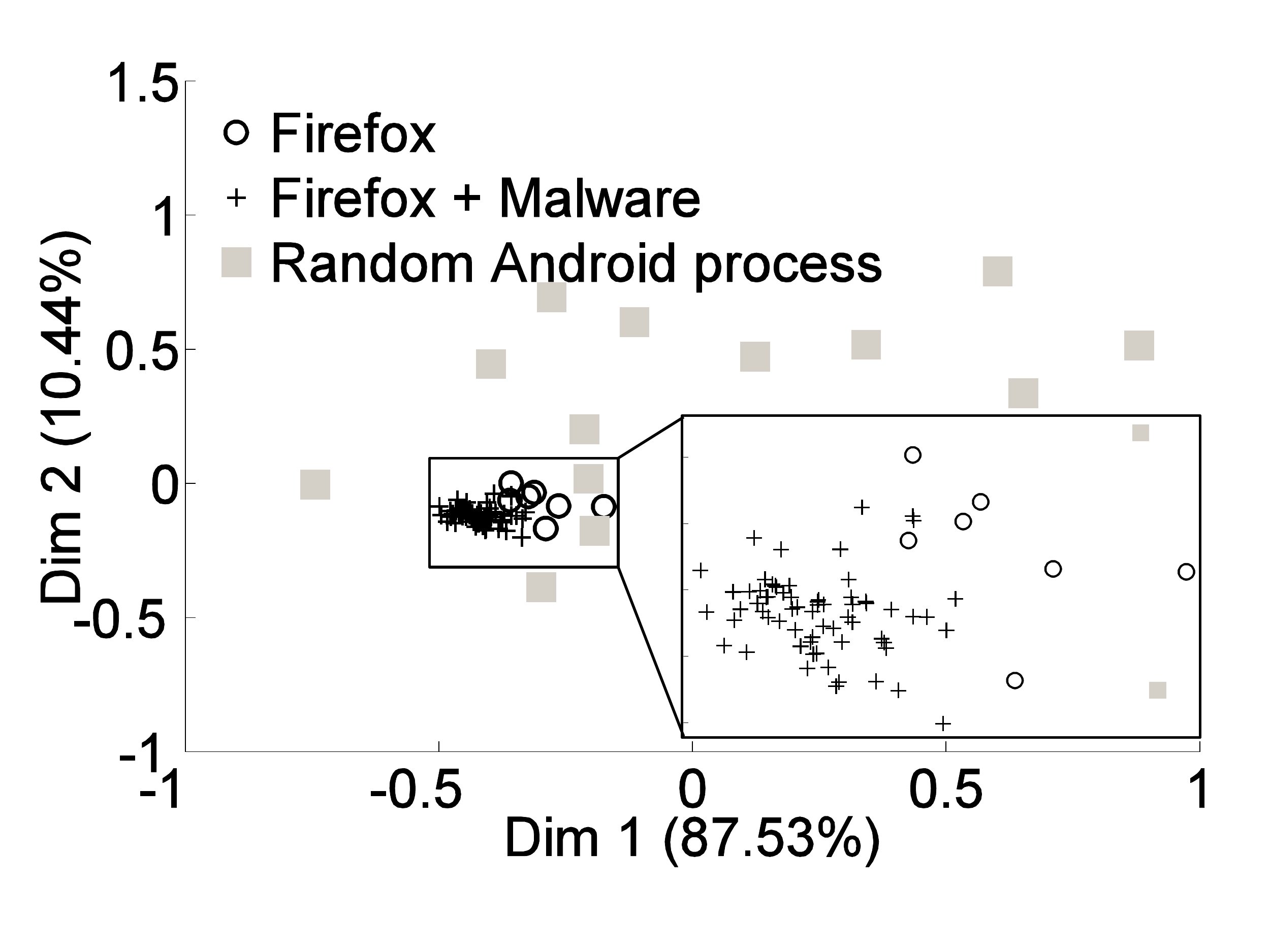}\par
\caption{Differential analysis of malware v. benignware.
The plot shows principal components of benign Firefox, 
Firefox with malware, and arbitrary Android apps. 
Malicious Firefox's traces are closer to Firefox than to random apps.
}
\label{fig:pca_benign_repack}
\end{figure}

Importantly, HMDs' value lies in being trustworthy and light-weight in
comparison to software based detectors, e.g., by running in an
enclave~\cite{sgx-1, sgx-2} secure against even user errors and kernel
rootkits~\cite{master-key}.  HMDs do {\em not} need to have 0\%
false positives and 100\% true positives---they only need to serve as an
effective filter for a global detector that can then use program
analysis~\cite{taintdroid,flowdroid} or network-based
algorithms~\cite{dash06} to build a robust global detector.  
We refer readers to Vasilomanolakis et
al.~\cite{cids-survey} for a survey on collaborative
malware detectors.

\subsection{Hardware-based Malware Detectors}
\label{sec:threat}

One line of HMD research focuses on {\em desktop} malware which has 
very different characteristics compared to mobile malware.
Ozsoy et al.~\cite{map} 
propose custom
hardware signals and hardware-accelerated classifiers and use
off-the-shelf desktop malware to evaluate their HMD with $\sim$90\% true positive
and 6\% false positive rates.
Tang et al.~\cite{anomaly-detector-columbia-raid} present
an anomaly detector for desktop malware and evaluate using 
2 benign programs and 3 exploits, achieving 99\% detection accuracy for less
than 1\% false positives.

To understand how Android malware is different, we 
compare 20 Windows malware samples (similar to ones in the studies above) to 20 benign
programs such as pdfviewer, calculator, filetransfer, resizer, screensaver,
etc.
We find that Windows malware executed an average of $\sim$60K system calls within 10
minutes v. only 2.5K for benignware. RegSetValue, the system call used to
modify Windows registry, is invoked 820 times by malware and only 72 times by
benignware.  Further, malware spawns 182 processes/threads on average
while benignware spawns fewer than 30.
Windows malware have historically targeted gaining control of the machine
whereas Android malware rarely attempt system-level exploits. Hence, 
mobile malware executions are far closer to benign executions. 
\mohit{Future: sentence about android malware syscalls}. We present our findings
about mobile malware in Section~\ref{sec:mobile-uniq} and quantify
these in Section~\ref{sec:results-anomaly}.


The closest related work to ours -- on HMDs for mobile malware -- is by Demme et
al.~\cite{simha-isca}, where the authors present a supervised learning
HMD that compares off-the-shelf Android malware to arbitrary benign apps,
yielding an 80:20 true positive to false positive ratio.  However, this
methodology of using off-the-shelf malware and comparing it to arbitary benign
apps is fallacious, as we discuss next.

\subsection{Pitfalls in Evaluating HMDs}
\label{sec:pitfalls}



One challenge in evaluating detectors is that 
malware developers can {\em adapt} their apps in
response to proposed defenses. For example, 
we have found that 
simply splitting a
payload into multiple software threads dramatically changes the malware's
performance-counter signature and training a signature-based HMD 
on the former execution yields
a very low probability of labeling the latter as malware.\todo{footnote to point
to longer tech report.}

Further, prior work analyzes malware samples categorized
by family-names like {\tt CruseWin} and {\tt AngryBirds-LeNa.C}---this does
not inform an analyst as to why a malware binary was (un)detectable.  Instead, we
propose that determining the robustness of a hardware-based malware detector
requires understanding 
{\em why} a particular malware sample was (un)detectable, to anticipate {\em
how} it can adapt, and then to create a malware benchmark suite to {\em
identify the operating range} of the detector.

A second challenge is that 
mobile malware
samples available online~\cite{dissect_malware_2012,contagiodump}, and used in prior
work, seldom execute `correctly' (Figure~\ref{fig:geinimi-crash}).
Malware often require older, vulnerable
versions of the mobile platform, they may target specific
geographical areas, include code to detect being executed inside an
emulator, wait for a (by now, dead) command-and-control server to
issue commands over the internet or through SMSs, or in many
cases, trigger malicious actions only in response to specific user
actions~\cite{obad,geinimi}. 
20\% of malware executions
in Demme et al's~\cite{simha-isca} experiments 
lasted less than one second and 56\% less than 10 seconds -- less time than it takes to 
steal 5 photos.
We posit that experiments should
establish that malware does execute its `payloads' -- such as stealing personal
information, tracking locations, sending premium SMSs etc -- instead of
executing a binary on a network-connected machine 
and assuming that payloads executed correctly~\cite{simha-isca,map}.

\begin{figure}[tbp]
   \centering
   \includegraphics[width=0.42\textwidth]{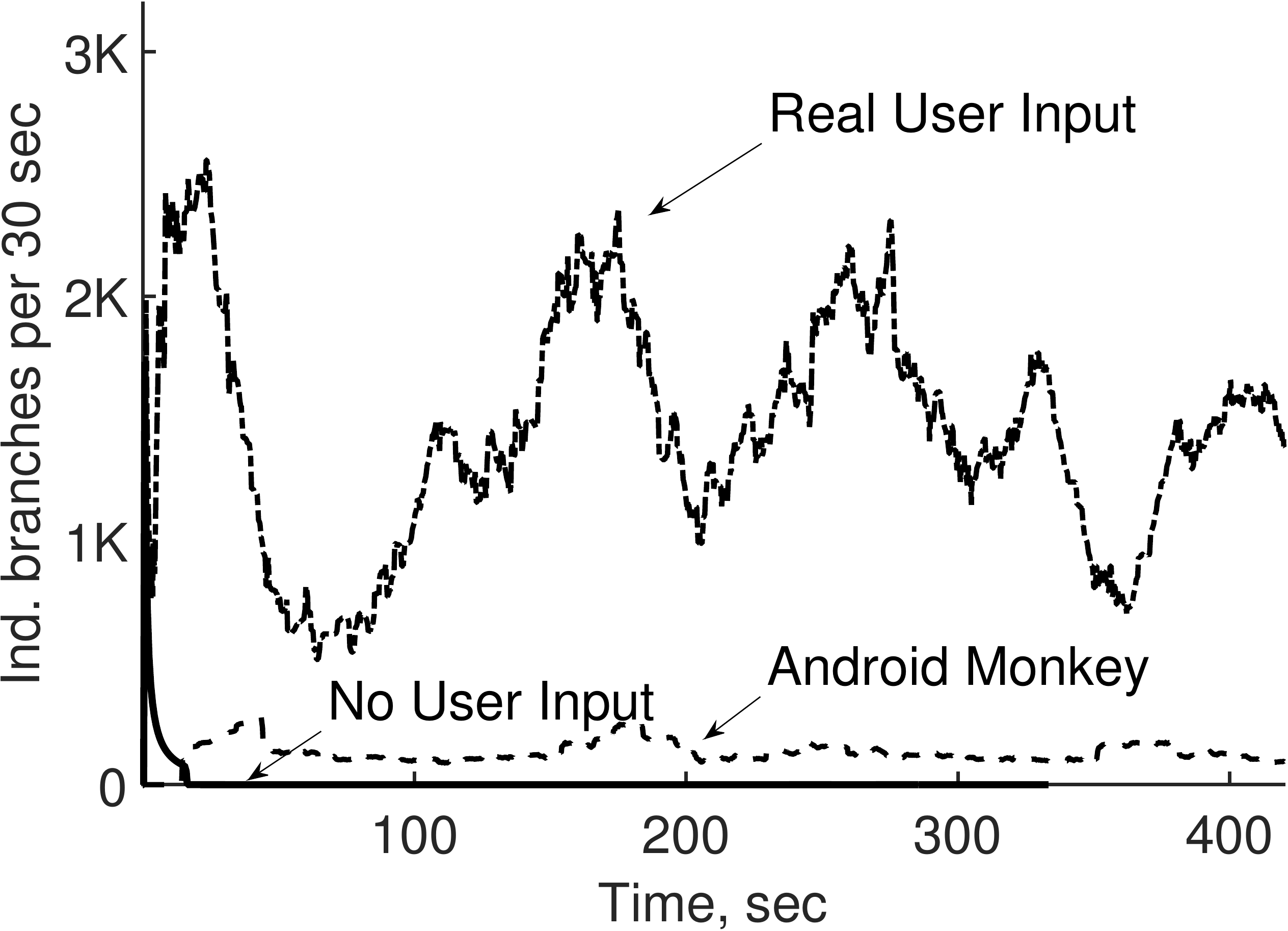}\par
   \caption{\todo{minipage with Fig 3}
   Real user inputs create hardware level activity, while providing no 
   input or using Android's input-generation tool (Monkey) creates a very small signal.}
\label{fig:ui_diversity}
\end{figure}

\ignore{\todo{malware taxonomy: explain how some samples were fixed.}}
 
A third challenge is to ensure appropriate differential analysis between benign
and malware executions.
Prior work~\cite{simha-isca} trains detectors on malware executions but
tests against {\em arbitrary} benign applications.  However,
Figure~\ref{fig:pca_benign_repack} shows that Firefox infected with malware
looks similar to Firefox itself and still very different from arbitrary Android
processes like {\tt netd}.  
Further, Figure~\ref{fig:ui_diversity} shows that driving Android applications
using real user-input has a major impact on the execution signals
compared to giving no input or using the Android `Monkey' app to generate random inputs.
Hence, we propose to 
test HMDs using malicious binaries against appropriate parent apps while both
apps are being driven using real user-inputs.

\noindent{\bf On Quantitative Comparison to Prior Evaluation Methods.} We have
shown in this section that prior `black-box' methods yield traces that do {\em
not} represent either malware or benignware executions. The prior method has
logical flaws -- as a result, 20\% of malware traces in~\cite{simha-isca} are
shorter than 1 second, and 56\% are <10s -- and we deliberately eschew further
quantitative comparisons with \name. Instead, our evaluation focuses on case
studies using \name to yield new insights into building effective HMDs.

\ignore{
\begin{figure*}[tbp]
\begin{minipage}[b]{0.33\linewidth}
   \centering
   \includegraphics[width=\textwidth]{figs/Motivation/AngryBirds/NEW_AngryBirdsSpace_MOTIVATION.pdf} \\
   Angry Birds game + click fraud 
   \label{fig:motivation-1}
\end{minipage}
\begin{minipage}[b]{0.33\linewidth}
   \centering
   \includegraphics[width=\textwidth]{figs/Motivation/Sana_Moca/NEW_MOCA_MOTIVATION.pdf} \\
    Sana medical app + SMS stealer
   \label{fig:motivation-2}
\end{minipage}
\begin{minipage}[b]{0.33\linewidth}
   \includegraphics[width=\textwidth]{figs/Motivation/TuneInRadio/NEW_RADIO_MOTIVATION.pdf} \\
   Internet radio + password cracker 
   \label{fig:motivation-3}
\end{minipage}
\caption{(Issues with black-list approach) Malware can adapt its computational signature to avoid
malware-signature based detection without reducing its own efficiency. Figures
show probability of having seen similar execution phases in the malware training set, 
so lower value implies less likelihood of being labeled as malware. The test malware (dark line)
looks even more unlikely to match malware training-set than benign programs (dashed line).
}
\label{fig:motivation}
\end{figure*}
}

%



\ignore{
\begin{figure}[tbp]
\centering
\includegraphics[width=0.45\textwidth]{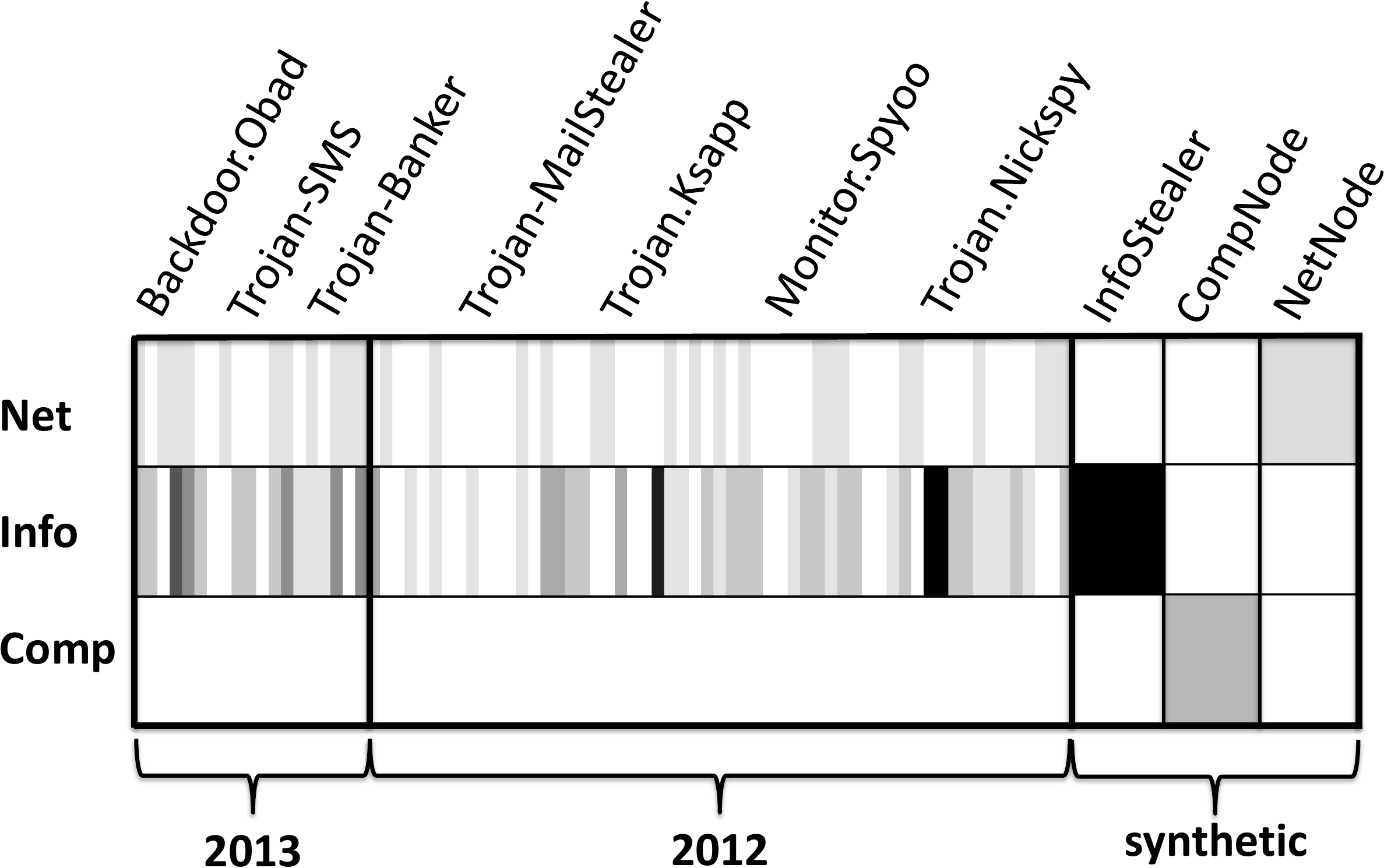}\par
\caption{Heat map}
\label{fig:heat_map}
\end{figure}
}
\ignore{
The best known technique of using performance counters to detect
malware~\cite{demme-counters} is similar to blacklisting approaches used in
anti-virus products today.  The supervised learning technique trains a
classifier on performance counter traces collected by executing instances of a
malware family (along with execution traces of arbitrary benign programs).  At
deployment time, performance counter traces from an actual machine are fed to
these classifiers (one per malware family) in order to identify malicious
activity.

The {\em Achilles' heel} of the current state-of-the-art technique is that
malware can adapt its signature to evade detection
(Figure~\ref{fig:motivation}). A malware developer has the option to modify
their code while retaining the malware's overall behavior. Hence, we experiment
with training a Markov model on one malware sample's performance counter
traces, and testing using performance counter traces from a different concrete
execution but the same behavioral payload.


Specifically, the new malware sample splits the original malicious workload
into multiple threads (compared to a single-threaded training set), and execute
the payload at a better performance. We use three benign applications -- one
that is user-driven, one that is compute-heavy, and a third that is network
intensive -- and repackage them with malware that (respectively) steals data,
fetches web pages to optimize the pages' search engine ratings, and executes a
password cracker.  We train the Markov model on 3 traces for each app (where
each trace is 96s -- 240s long), and test it on the modified payload. 

We defer the Markov model discussion to Section~\ref{sec:arch} and plot the
results in Figure~\ref{fig:motivation}. Figure~\ref{fig:motivation} shows the
probability that the testing trace is similar to the training set over time.
Our experiment shows that in all three repackaged apps, the malware can find an
alternate execution profile that is far removed from its previous signature
(over two standard deviations from the training set). In fact, we find that the
new malware (dark line) is even more unlikely to be labeled as similar to the
training set than the underlying benign program (the dashed line). This evasion
shows that an alternative approach is required to detect new malware whose
concrete executions can vary even if behavior remains the same.
}


\section{Malware Taxonomy}
\label{sec:malware-gen}

%

The first major component of \name generates a diverse population 
of malicious apps.
To do so, we first introduce a taxonomy of high-level malware behaviors, 
and then use it to create a set of representative malware whose 
hardware signals have been explicitly diversified.




\ignore{
\begin{table}[ht]
\caption{Behavioral classification} 
\centering 
\begin{tabular}{c c c c} 
\hline\hline 
Year & Info. Theft & Networked Nodes & Compute Nodes \\ [0.5ex] 
\hline 
2012 & 68.8\% & 69.5\% & 0 \\ [1ex] 
2013 & 38.2\% & 60.6\% & 0 \\ 
\hline 
\end{tabular} 
\label{table:highlevel_classification} 
\end{table} 
}

\begin{figure}[tbp]
   \centering
   \includegraphics[width=0.45\textwidth]{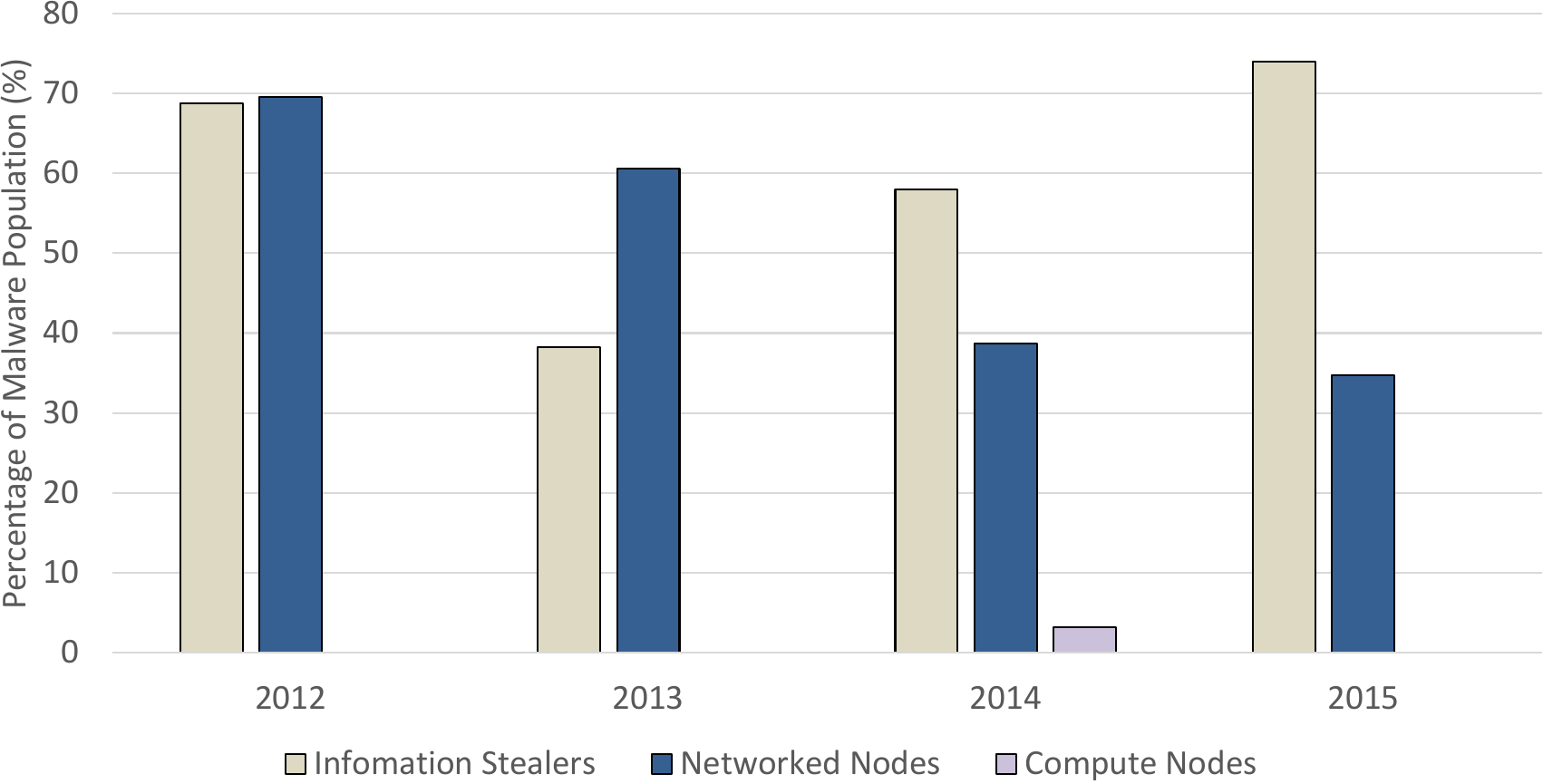}
   \caption{Malware behaviors observed in a 126-family 229-sample Android malware
set from Contagio minidump.  Most malware steals data or carries out network
fraud.  However, samples that use phones as compute nodes, e.g., to crack
passwords or mine bitcoins,  have been reported in 2014.}
\label{fig:malware-hilevel}
\end{figure}

\begin{figure}[tbp]
   \centering
   \includegraphics[width=0.45\textwidth]{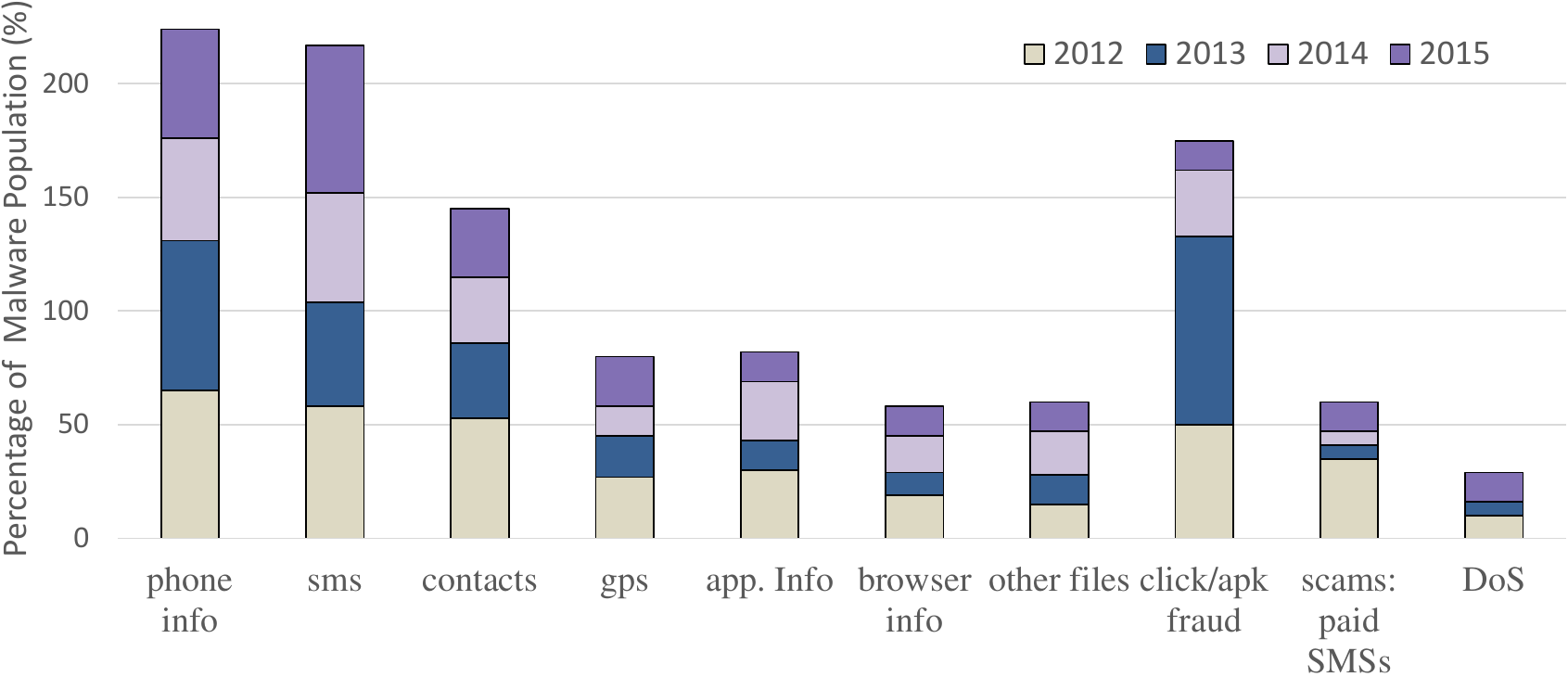}
\caption{Examples of malware behaviors and their contribution to the malware
dataset.}
\label{fig:malware-breakdown}
\end{figure}

Figures~\ref{fig:malware-hilevel} and ~\ref{fig:malware-breakdown} 
show our manual classification of malware into high level behaviors. We studied 
53 malware families from 2012, 19 from 2013, 31 from 2014 and 23 from 2015
--  a total of 229 malware samples in 126 families -- downloaded from
public malware repositories~\cite{contagiodump,malware.lu,virusshare.com}. 
Our classification's goal is to identify 
orthogonal atomic actions
and to determine 
concrete values for these actions (e.g., amount and rate of data stolen). 

To classify malware, we disassembled the binaries (APKs on Android) and
executed them on both an Android development board and the Android emulator to
monitor: permissions requested by the application, middleware-level events
(such as the launch of Intents and Services), system calls, network
traffic, and descriptions of malware samples from the malware repositories.
We describe our findings below.

\subsection{Unique Aspects of Mobile Malware}
\label{sec:mobile-uniq}


Our key insight is that instead of trying to detect conventional root {\em
exploits}~\cite{rage-against-the-cage, gingerbreak, exploid}, we propose to
detect malicious {\em payloads}. Here, payloads refer to code that achieves the
malware developers' goals, such as sending premium SMSs, stealing device IDs or
SMSs, etc.  We observed root exploits in only 10 of 143 samples in 2012 and 3 of 32
samples in 2013 -- we now take a closer look at the attack vectors
mobile malware rely on.


Mobile malware can successfully execute payloads due to vulnerable third-party
libraries. In one instance that affected hundreds of millions of
users, a ``vuln-aggressive'' ad-library had a deliberate flaw
that led to downloaded files being executed as code~\cite{ad-vulna}.  Webviews, that enable
Android apps to include HTML/javascript components, are another major source of
vulnerabilities~\cite{webview-attacks} that allows payloads to be dropped to a
device.  Apps with this vuln-aggressive library or Webviews are otherwise benign
and can be downloaded from app stores as developer signed binaries, only to be
compromised when in use. 

In other cases, errors by an app's benign developers themselves can lead to
malicious payloads being executed.  Misconfigured databases even in popular
apps like Evernote~\cite{evernote} and AppLocker~\cite{applocker} (a secure
data storage app) were vulnerable to malicious apps on the device simply
reading out data from sensitive databases.  In such cases, the malicious app
could be an otherwise harmless wallpaper app that constructs an `Intent' (a
message) to AppLocker's database at run-time and exfiltrates data if successful.

User errors are another cause for malware payloads executing successfully at
run-time. Malicious apps read data from an online server, use it to construct a
user prompt at run-time, and thus request sensitive permissions such as access
to SMSs or microphone. Users often accept such
requests~\cite{user-error-permissions} and once authorized, apps can siphon off
{\em all} SMSs or conduct persistent surveillance attacks~\cite{RAT}.

Worst of all, even the platform (Android) code can have severe vulnerabilities
that doesn't require a conventional exploit. For example, the Master Key
vulnerability~\cite{master-key} simply involved an error in how Android
resolves a hash collision due to resource-names in a binary at install time v.
execution time. By packing the binary with a malicious payload such that the
install time check passes but the execution time loader picks the other
malicious payload, attackers could distribute their payloads through signed
apps in official app-stores.

\parbold{Finding: analyze payloads instead of exploits.}
Based on the above findings, we 
conclude that while there are many routes to getting a payload
to execute as part of a benign app, executing the payload is mandatory for
malware to win.  Hence our proposed detectors seek to {\em distinguish malicious payloads from
benign app executions}.  The challenge of detecting payloads is that payloads can
look very similar to benign app's functionality. For example, if a previously
harmless game AngryBirds starts to comb through a database, can we distinguish
whether it is reading a user's gaming history (harmless) or a user's password
database (attack) using only hardware signals.  

In this sense, our problem is a general and more complicated version
of detecting exploits -- whether Internet Explorer or Acrobat PDF is under
a return-oriented programming attack (ROP) followed by Stage 1
shell codes (as considered by Tang et al~\cite{anomaly-detector-columbia-raid}).


\ignore{TODO: Repackage into
a favorable app that already requests permissions to access sensitive resources and
internet access. Since resources
like SD card, camera, internet, etc can be accessed simply by asking for a
permission, we believe there is little motivation for malware to rely on root
exploits.  TODO: Add webviews and ad library vulnerabilities.
This trend motivates our high level approach of identifying malware
based on their behaviors instead of looking for signatures of specific
exploits.
}
\ignore{
A common feature of most existing Android malware is that they rely heavily on
social engineering techniques instead of root exploits~\cite{trendmicro}.
Developers of malicious applications (``apps'') include them into benign apps,
especially into popular paid apps, and distribute on a third party markets for
free. This method is called "repackaging"~\cite{dissect_malware_2012} and
several studies have attempted to identify repackaged applications in Android
markets~\cite{ling-repackaged}. Another popular method of distributing Android
malware is to create ``renamed'' apps: a fully malicious app uses a name of a
legitimate app. For example, a fake anti-virus app.  When a user installs such
an app, it quickly executes its malicious payoad, shows an error message
directing the user to a legitimate app and quits.
}


\ignore{Also, we wanted to analyze how Android malware has evolved since the time
~\cite{dissect_malware_2012} was published.  -- If time permits, we will
compare with NCSU study.  }

\subsection{Behavioral Taxonomy of Mobile Malware} 
\label{sec:taxonomy}

At a high level we assigned every malicious payload to one or more of three
behaviors: {\em information stealers, networked nodes}, and {\em compute
nodes} (Figure~\ref{fig:malware-breakdown}).  

Information stealers look for sensitive data and upload it to the server.
User-specific sensitive data includes contacts, SMSs, emails, photos, videos, and
application specific data such as browser history and usernames, among others.
Device-specific sensitive data includes identifiers -- IMEI, IMSI, ISDN -- and
hardware and network information.  The volume of data ranges from photos and
videos at the high end (stolen either from the SD card or recorded via a
surveillance app) to SMSs and device IDs on the low end. 


The second category of malicious apps requires compromised devices to act as
nodes in a network (e.g., a botnet).  Networked nodes can send SMSs to premium
numbers and block the owner of the phone from receiving a payment confirmation.
Malware can also download files such as other applications in order to raise
the ranking of a particular malicious app.  Click fraud apps click on a
specific web links to optimize search engine results for a target.


Given the advances in mobile processors, we anticipated a new category of
malware that would use mobile devices as compute nodes. For instance, mobile
counterparts of desktop malware that runs password crackers or bitcoin miners
on compromised machines. This was confirmed by recent malware samples whose
payload was to mine cryptocurrencies~\cite{mobile-bitcoin-miner}.  We did not
observe Bitcoin miners until mid-2014 (when we conducted our survey) and 
used a password cracker as a compute-oriented malware payload. The cracker's
task is to recover sensitive passwords by making a
guess, compute the guess' cryptographic hash, and compare each hash against
a sensitive database of hashed passwords.

\begin{figure}[tbp]
   \centering
   \includegraphics[width=0.45\textwidth]{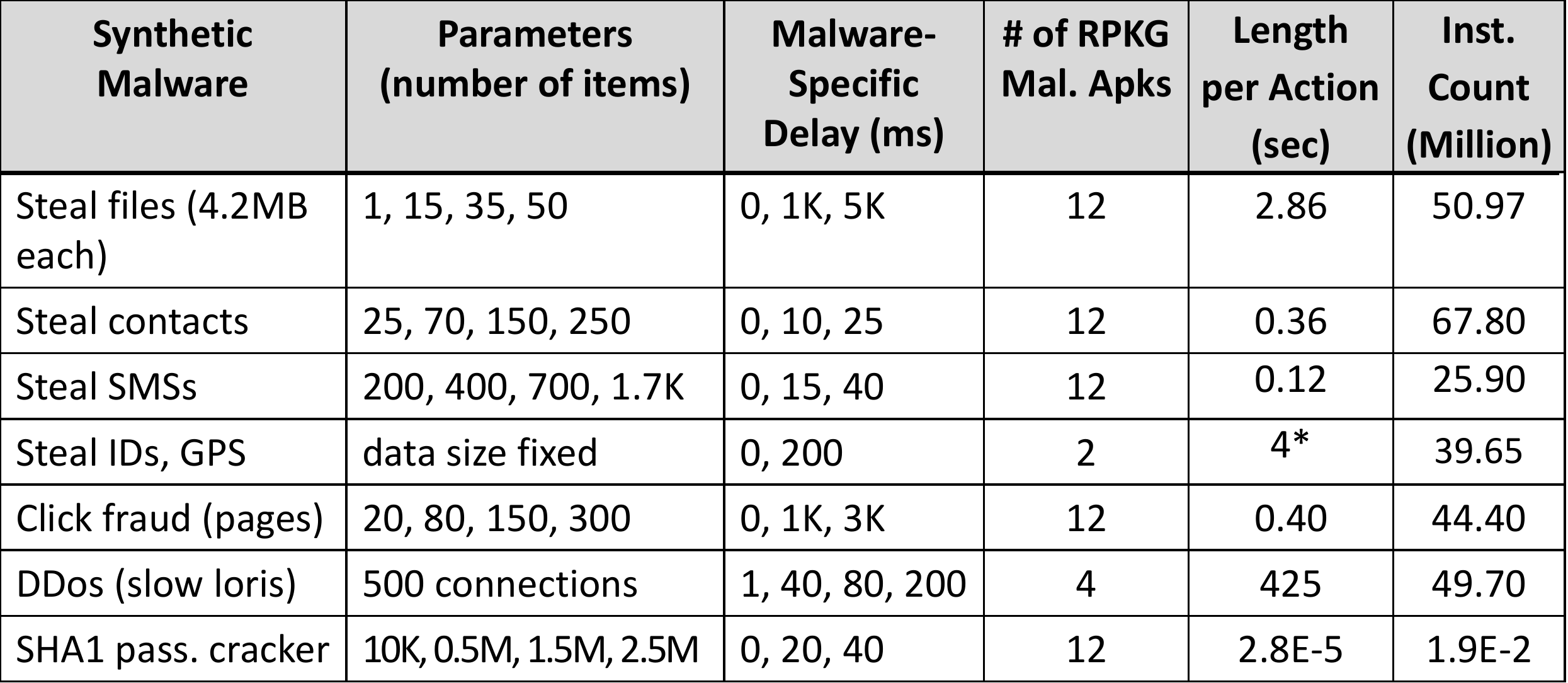}
\caption{
Malware payloads: 4 info stealers, 2 networked nodes, and 1 compute node. 
These settings represent a small but computationally diverse subset of malware behaviors. Interestingly, small software actions have large hardware footprints. 
}
\label{fig:synthetic-malware}
\end{figure}

{\bf Finding: Software-level actions are surprisingly long in hardware.}
Figure~\ref{fig:synthetic-malware} shows the specifics of each malware behavior
we currently include in \name. 
Interestingly, {\em atomic} malware payload actions take significant amount of
time at the hardware level for several payloads -- e.g., stealing even one SMS
or a Contact requires 0.12s to 0.36s on average. These constants inform the design of
our performance counter sampling durations and machine learning models in Section~\ref{sec:impl}.
The last two columns in
Figure~\ref{fig:synthetic-malware} show the average length of an atomic action
in the malware payload (not counting 
delays such as being scheduled out by the operating system), 
and the instruction count per action (e.g. stealing 1
photo/contact/SMS, clicking on 1 webpage in click fraud, opening 500
connections and keeping them alive in a DDoS attack, generating 1 string and
computing its hash using SHA1).\ignore{data collected over how many runs?}

%

\ignore{

After identifying a high-level structure for malicious applications, we dug
deeper in order to understand the underlying black market model and set
appropriate parameters for the malware categories.
\ignore{We pose a question why
people collect someone else's personal data, because we want to estimate how
important information stealing in mobile world is and will be in the future.
}

Information stealing mobile applications spy on a person or group of people by
surreptitiously harvesting their emails, call logs, text messages, contact
lists, and even camera and microphone inputs.  Some companies officially bundle
such functionality into spywares~\cite{copy9}.  Such data collection represents
the high end in terms of volume of data stolen.  Note that benign applications
may also collect this personal information to better serve their users -- i.e., a
digital assistant like Google Now to learn user preferences, or by advertising
libraries for marketing purposes. Our task is to identify when an application
uncharacteristically accesses these data sources.

On the other hand, stealing sensitive device-specific IDs does not affect
the privacy of real people. On the black market, stolen device IDs are used to
legalize stolen physical mobile devices.  A network provider may ban a device
with a particular ID to connect to the cell network after receiving a report
that a device has been stolen.  However, criminals replace IDs of a stolen
device with IDs of a legitimate device that can be bought on the black market. 
Device IDs are thus valuable and represent the low end in terms of amount of 
data that is exfiltrated by a malicious app.

Networked nodes involve malware that uses a device primarily as a client on the
network, instead of as a direct target.  One popular use case is to send SMSs
to premium numbers and block the owner of the phone from receiving a payment
confirmation.  Programs also download files such as other applications in order
to raise the ranking of a particular malicious app even in an official app
store.  Click fraud apps click on a specific web links to optimize search
engine results for a target or to cause a victim to bleed money.  Finally,
although the NCSU dataset~\cite{survey_malware_2011} found that mobile phones
present challenges for DDoS attackers because of low bandwidth and low battery
capacity, network bandwidth has improved substantially since then, and we
observe a few modest attempts to build DDoS attackers.



We have not observed malware that uses mobile devices as compute nodes, but we
expect that such applications, such as password crackers, will emerge
when mobile device performance improves with better CPUs and more memory. 
}

\ignore{
\begin{table}[ht] 
\caption{Info Stealing} 
\centering 
\begin{tabular}{c c c} 
\hline\hline 
subcategory & 2013 & 2012 \\
\hline 
 phone info & 65\% & 66\% \\ 
 sms        & 58\% & 46\% \\ 
 contacts   & 53\% & 33\% \\ 
 apps info  & 30\% & 13\% \\ 
 GPS        & 27\% & 18\% \\ 
 email      & 19\% & 8\% \\ 
 browser    & 19\% & 10\% \\ 
 files      & 15\% & 13\% \\ [1ex] 
\hline 
\end{tabular} 
\label{table:info_stealers} 
\end{table} 

\begin{table}[ht] 
\caption{Networked Node} 
\centering 
\begin{tabular}{c c c} 
\hline\hline 
subcategory & 2013 & 2012 \\
\hline 
 download files & 50\% & 83\% \\ 
 scam 	     & 35\% & 6\% \\ 
 DDoS        & 10\% & 6\% \\ 
 click fraud & 5\%  & 0\% \\ [1ex] 
\hline 
\end{tabular} 
\label{table:net_node} 
\end{table} 
}

\ignore{
\begin{table}[ht] 
\caption{Synthetic malware generator} 
\centering 
\begin{tabular}{c c c} 
\hline\hline 
malware basic blocks & ~ & ~ \\
\hline 
Info stealers & ~ & ~ \\ 
\hline 
 sms 		& online & offline \\ 
 phone info 	& ~ & ~ \\ 
 contacts       & ~ & ~ \\ 
 GPS 		& ~ & ~ \\
 browser info 	& history & bookmarks \\ 
 SD card 	& files & dir content \\ 
 apps info 	& ~ & ~ \\ 
\hline 
Net node & ~ & ~ \\ 
\hline 
 download 	& file(s) & ~ \\ 
 DDoS 		& IPs & ~ \\ 
\hline 
Comput. node & ~ & ~ \\ 
\hline 
matrix mult 	& A & B \\ [1ex] 
\hline 

\end{tabular} 
\label{table:mal_gen} 
\end{table} 
}

\ignore{
After identifying a common set of malicious behaviors, we created a tool that
can model both the malware that we found in the wild and malware that can
mutate its computations and yet achieve the same ends.  Our framework helps us
to study the limits of using computational signatures to detect malware.
}

\subsection{Constructing Malware Binaries} 
\label{sec:diverse-malware}

We now describe the steps required to create a realistic malware binary.
%
Malware activation can be chosen from being triggered at
boot-time, when the repackaged app starts, as a response to user activity, or
based on commands sent over TCP by a remote command and control (C\&C) server.  In all
cases, malware communicates back to the C\&C server to transfer stolen data or
compute results.  \name's configuration parameters also specify 
network-level
intensity of malware payload in terms of data packet sizes and interpacket
delays, and device-level intensity in terms of execution progress (in terms of
malware-specific atomic functions completed).  We chose concrete parameters for
malicious payload based on an empirical study of mobile malware as well as
information about benign mobile devices~\cite{morpheus-hasp}. 

\begin{figure}[tbp]
   \centering
   \includegraphics[width=0.45\textwidth]{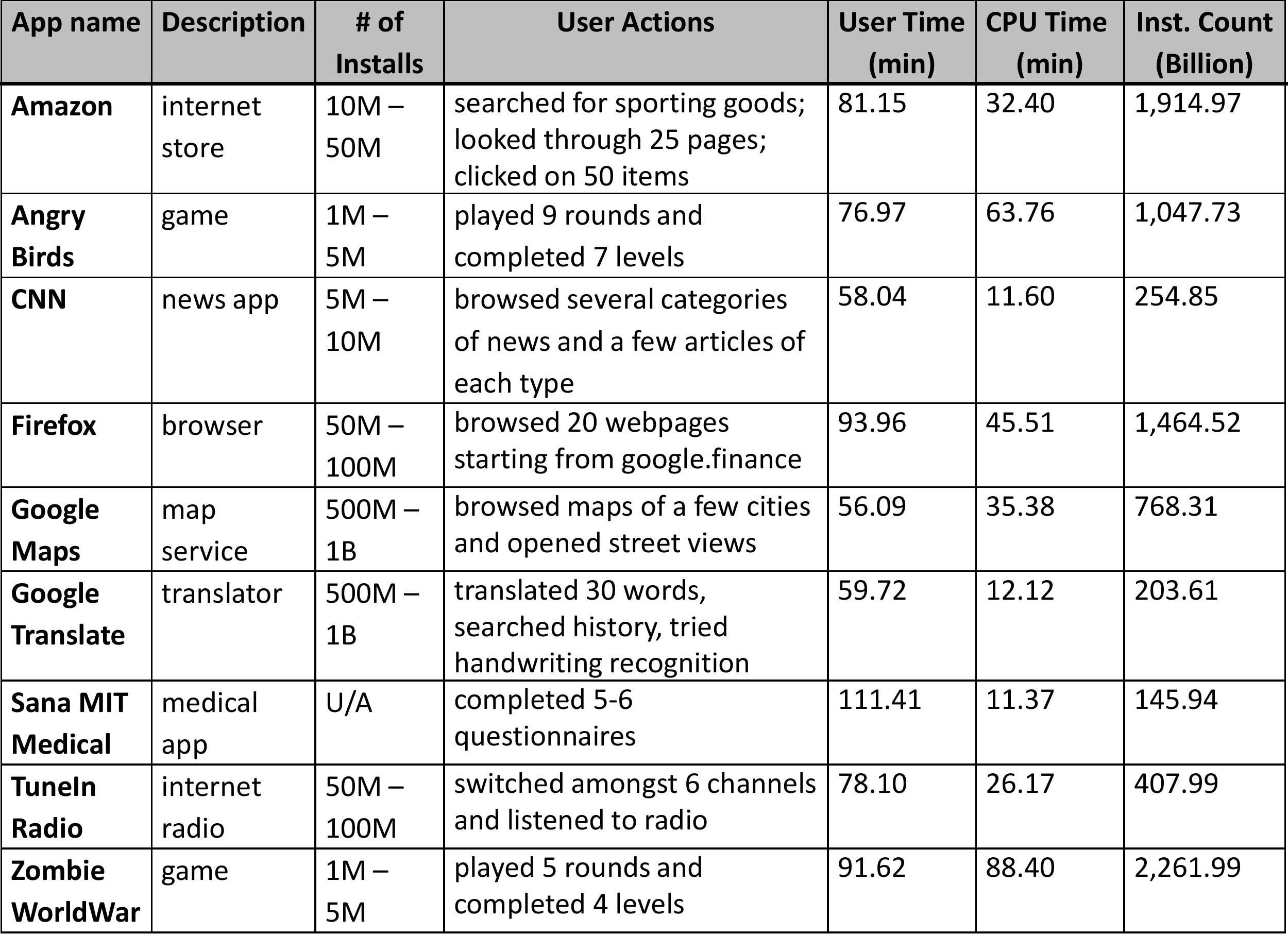}\par
\caption{Real user inputs on benign apps, with per app 
traces up to $\sim$2 hours and $\sim$2 trillion instructions. 
We choose complex apps and include a mix of compute (games), user-driven (browsers, medical app), and network-centric (radio) apps.}
\label{fig:baseline-apps}
\end{figure}

The generated malware has a top-level dispatcher service that serves as an
entry point to the malicious program; it parses the supplied configuration
file, launches the remaining services at random times, and configures them. Malicious
services can run simultaneously or sequentially depending on the configuration parameter.
In some cases, the service
that executes a particular malicious activity can serve as an additional dispatcher. For
example, the service executing click fraud spawns a few Java threads to avoid
blocking on network accesses. Every spawned thread is provided with a list of
URLs that it must access. Besides Android services, we register a listener to
intercept sensitive incoming SMS messages, forward them to C\&C server, and remove
them from the phone if needed. This listener simulates bank Trojans that remove
confirmation or two-factor authentication messages sent by a bank to a customer.  
%


Most professional apps are obfuscated using Proguard~\cite{GoogleProGuard}
to deter plagiarism. Proguard shrinks and optimizes binaries, and
additionally obfuscates them by renaming classes, fields, and methods with
obscure names.  We applied Proguard to the malware payloads (even when we did
not use reflection and encryption) to make the payloads look like real
applications.

After a malware payload is created, it must be repackaged into a baseline app.
Repackaging malware into a baseline app involves
disassembling the app (using {\tt apktool}), and adding information about
new components and their interfaces in the application's Manifest file.
We then insert
code into the Main activity to start the top-level malware dispatcher service 
(whose activation trigger is configurable), and add malicious code and data files into
the apk. We then reassemble the decompiled app using {\tt apktool}. If
code insertion has been done correctly, apktool produces a new Android app,
which must be signed by {\tt jarsigner} before deployment on a real device.

%


\ignore{
: information theft, networked node, and computational malware. We then
generate concrete instances of malware based on this configuration file by
injecting malicious payload into benign Android applications.  We also model
different activation mechanisms. For instance, malware can be triggered
automatically at boot time or when the baseline application starts, as a
response to user activity, or based on commands from a remote command and
control server (C\&C server) that guides malware execution.  In all cases, the
malware communicates with a remote server to transfer back data and/or results
of completed actions.}

\ignore{For information theft, we can set
the amount of data to be stolen from the SD card, contacts list, text messages,
and other sensitive user databases. Table~\ref{table:knobs} shows the specific
knobs we have implemented for each malware category. 
thus the injected malware resembles a set of malware basic blocks, whose execution is
controlled either by a static configuration file or by a remote server.}

\ignore{
Intensity of malicious behavior is controlled on two levels: network level and
local level. Network level intensity has two parameters: the size of each data
chunk transmitted and the delay between two consequtive transmissions. On the
local level, parameters are specific to each malware category.  In general,
malicious code is divided into small atomic parts that can be configurably
executed as multiple threads (or background services in Android) and a delay is
inserted between these atomic parts of the code. For example, the malware basic
block that carries out DDoS attack has a few local intensity paramters: delay
between requests, amount of data sent in each request, it also has a set of IP
addresses of the hosts under attack as well as the number of requests or the
longitude of the attack.  
}
\ignore{Malware categories and a summary of their specific
parameters are presented in the table ~\ref{mal_gen}. 
}


\ignore{Information stealing malware can be online and  analyze
incoming messages, forward some of them to the attacker, and remove some from
the device. This method is also used by real malware to intercept and delete
authentication messages coming from financial institutions. The {\em offline}
method of stealing text messages means uploading all or some of the currently
present on the device messages to the attacker's server. When stealing browser
data, malware can transmit to the attacker's server browser history and/or
bookmarks.  Also, our malicious payload can steal the entire files or the
content of particular folders on the SD card.  Networked nodes are represented
by click-fraud (downloading web-pages) and DDoSing of network hosts.
Computationally intensive workload is represented using a password cracking
tool that tries dictionary attacks on a password file.
}



\ignore{We next describe our mechanisms to construct computational signatures of benign
programs before putting the signatures to test using a diverse array of
malware.}


\ignore{Based on the comprehensive analysis of the existing Android malware and
of the underlying black market economy, we chose the most widespread malware
types that are prevalent today and very unlikely to become extinct in the near
future. The goal of our research is not only to develop a lightweight detection
system, but also to study how malware may adapt to such detection schemas. The
only way for a malware to escape detection is to hide itself within a benign
activity. In order to study such effect we built a malware generator with
numerous parameters controlling the intensity of the malicious behavior.  }

\section{Real User-driven Execution}
\label{sec:impl}
\begin{figure*}[t]
\begin{minipage}[tbp]{0.33\linewidth}
   \includegraphics[width=\textwidth]{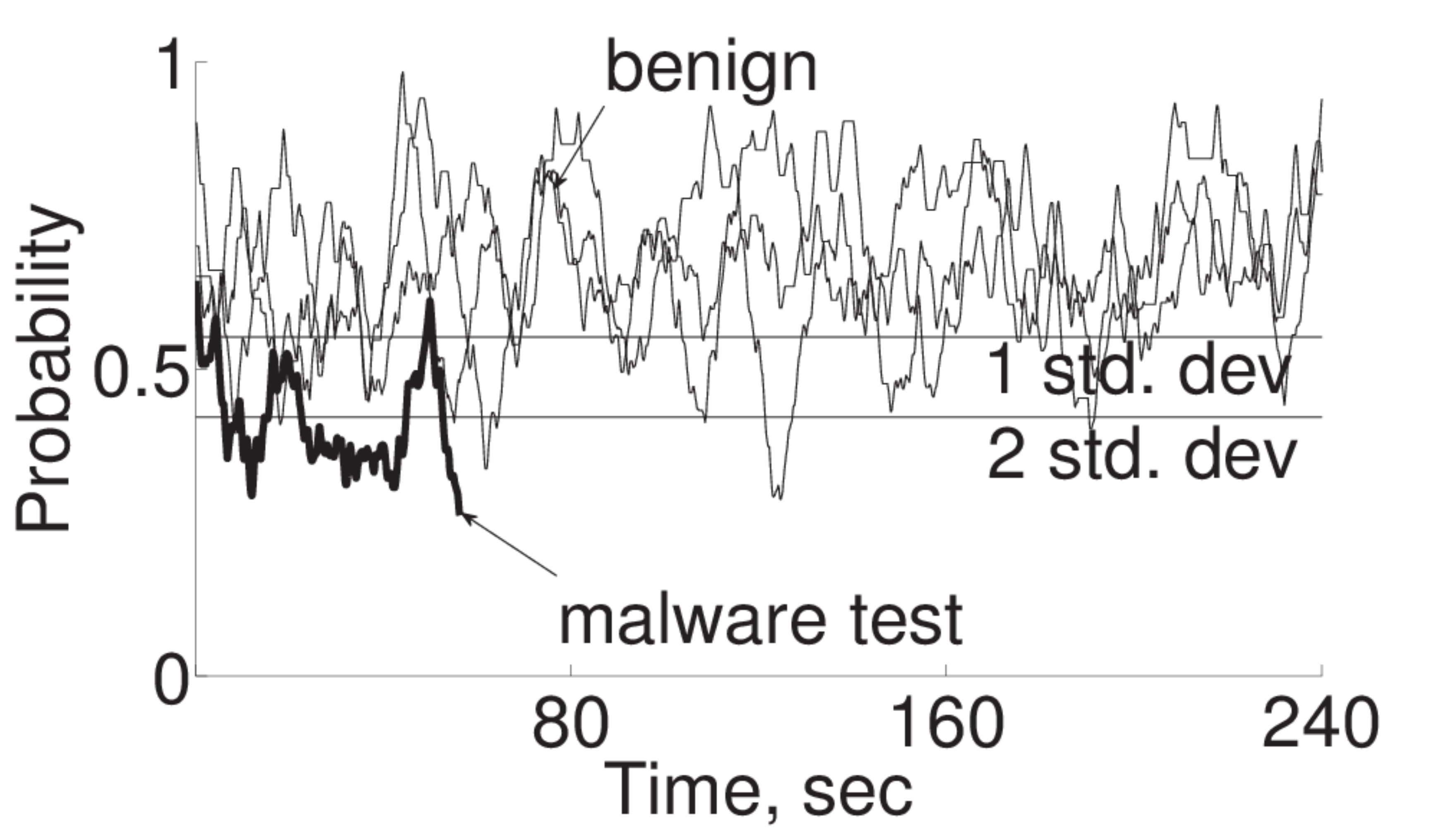}\par
   \label{fig:ab-1}
\end{minipage}
\begin{minipage}[tbp]{0.33\linewidth}
   \includegraphics[width=\textwidth]{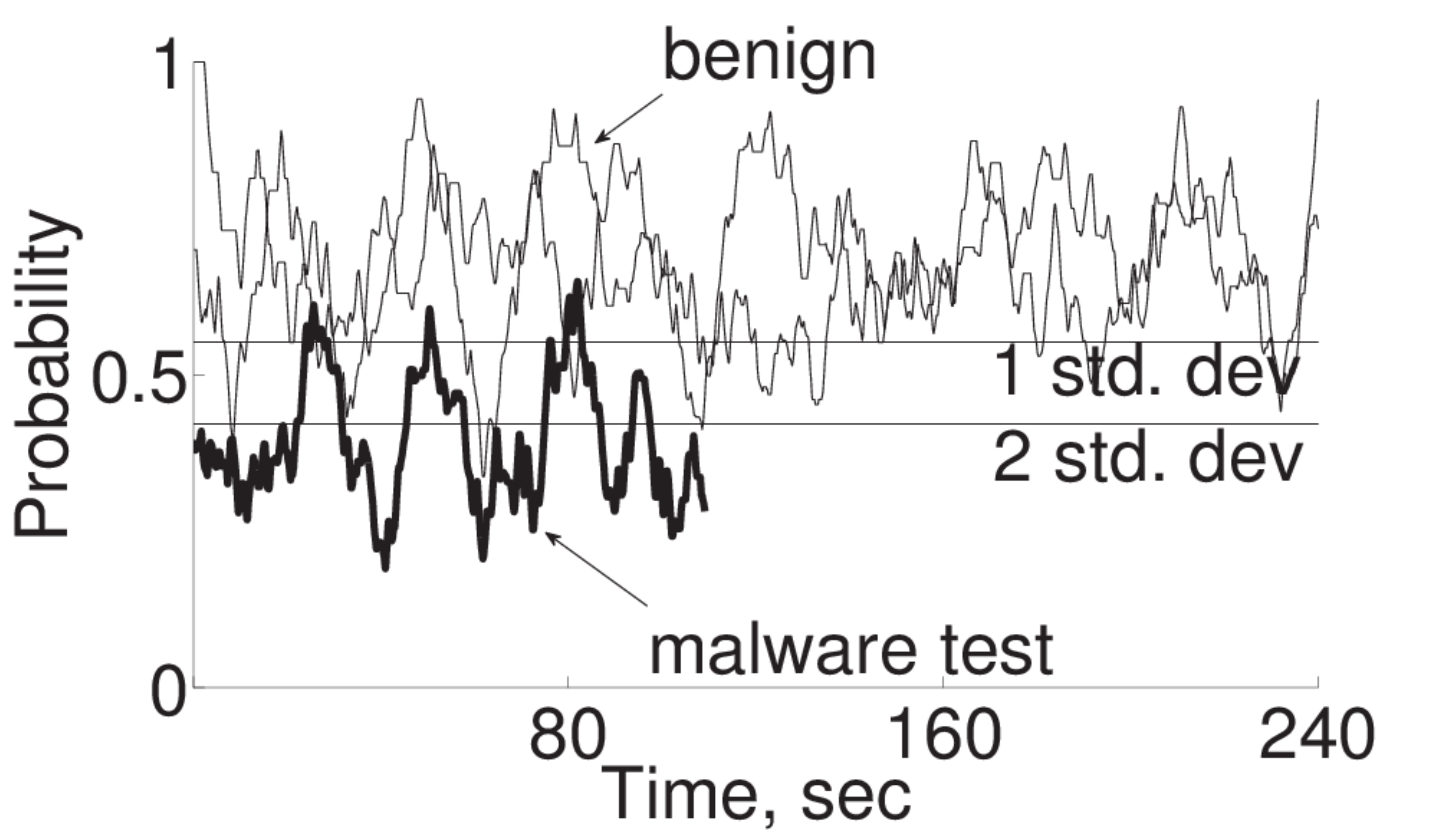}\par
   \label{fig:ab-2}
\end{minipage}
\begin{minipage}[tbp]{0.33\linewidth}
   \includegraphics[width=\textwidth]{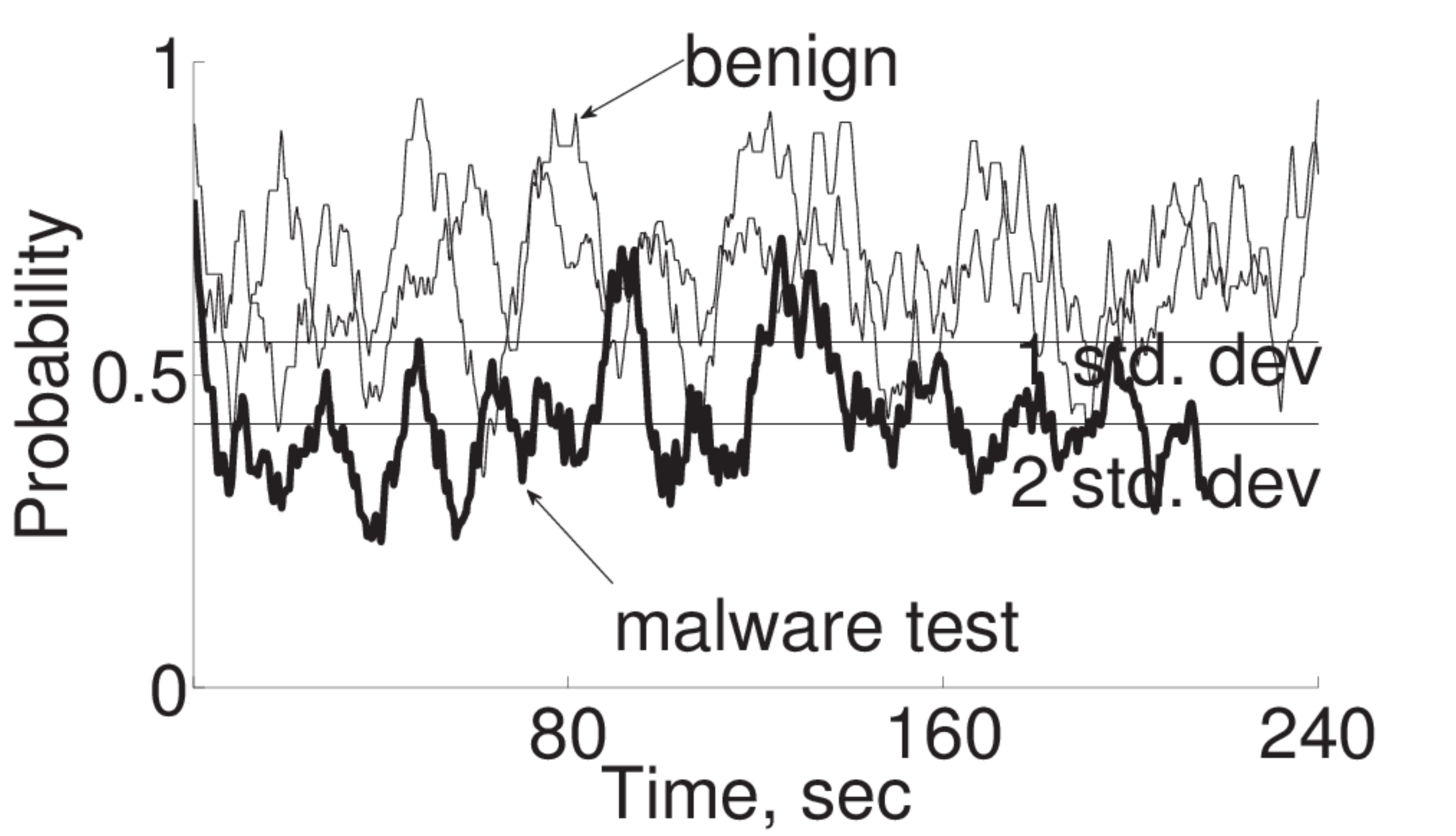}\par
   \label{fig:ab-3}
\end{minipage}
\caption{
HMD results for
Angry Birds with click fraud operating at three (increasing) intensities. 
Since HMD is trained on benign AngryBirds, 
a low dark-line shows that the HMD detects malware as a low probability
state.
}
\label{fig:ab}
\end{figure*}

Armed with a computationally diverse malware suite, we now select
a similarly diverse suite of benign apps, drive them with long, real, user
inputs, and extract hardware signals from them.  Figure~\ref{fig:baseline-apps}
shows the apps that we selected -- notice that we drive real user-level
functionality instead of random inputs.



\subsection{Benign Apps}
\label{sec:apps}
Our main goal is to choose applications that represent popular usage, and that
require permissions to access resources like SD card and internet connectivity.
This ensures that the applications are interesting targets for malware.
Further, we ensure that the apps cover a mix of compute (games),
user driven (medical app, news), and network (radio) behaviors,
diversifying the high-level use cases for apps in the benignware suite.
Our chosen app set includes native (C/C++/assembly), Android (Dalvik
instructions), and web-based functionality, varying the execution environment of our benign app pool. 
In our evaluation (Section~\ref{sec:results}), we confirm that this high-level
diversity does indeed translate into diverse hardware-level signals.

%


\subsection{User Inputs}
\label{sec:apps-inputs}
For each benign application, we created a workload that represents common
users' behavior according to statistics available online. For example, when
exercising Firefox, we visited popular websites listed on alexa.com. Automating
this is simple. For Angry
Birds, we recorded a user playing the game for multiple rounds and successfully
completing several levels.  For the medical diagnostics app (Sana), we record
users completing several questionnaires, where each questionnaire requires
stateful interactions spread over several screens.  Such deep exploration of
real apps is far beyond the capability of not only the default UI testing tool
in Android (Monkey~\cite{monkey}), but also state of the art in input
generation research~\cite{mayur-naik-gatech}.  Without such deep
exploration of benign apps, the apps' hardware traces will reflect only a
dormant app and cause the malware signals to stand out at test time but not in
a deployed system. 

For each benign app, we collect 6 user-level sessions (each 5--11 min long) and
use a heavily modified Android Reran~\cite{reran} to record and replay 4 of these sessions with
random delays added between recorded actions (while ensuring correct execution
of the app).  These 10 user-level traces per app generate 56--111 minutes of
performance counter traces across all apps.  

Each benign app is then repackaged with 66 different payloads to create 9
$\times$ 66 malware samples.  To collect performance counter traces, we replay
one of the app's user-level traces
and extract 5--11 minutes long performance counter traces for {\em each}
malware sample. 

Figure~\ref{fig:baseline-apps} shows some interesting trends in benign traces.  
While Sana commits 145 Billion instructions in 111 minutes, Zombie WorldWar
commits 2,261 Billion instructions in 91 minutes -- clearly, Sana is much more
user-bound while Zombie WorldWar is compute-heavy. CNN and Angry Birds are 
similar to Zombie WorldWar, where TuneIn Radio lies between Sana and Zombie 
WorldWar in instructions committed.


{\bf Finding: HMDs have to be application-specific.}
Interestingly, as we show in our evaluation (Section~\ref{sec:results}), the
compute intensity of CNN and Zombie WorldWar results in them having the worst
detection rates among all the apps in our suite. On the other hand, even though
TuneIn Radio is more intense than Sana, TuneIn Radio exposes malware better. We
find that this is because the Radio has more regular behavior while Sana
executes in short, sharp bursts. \name's realistic replay infrastructure and user-input
traces are key to producing these insights into HMDs' performance in a realistic setting.








\subsection{Extracting Hardware Signals}
\label{sec:signal-collection}

We now describe our measurement setup for precise reproducibility.  The
measurement setup requires careful setup and correctness checks since it is
difficult to replay real user inputs to the end once delays and 
malware payloads are added.  


\noindent{{\bf Devices.} Our experimental setup consists of an Android
development board connected to a desktop machine via USB, which in turn stores
data on a server for data processing and construction of ML models.  The
desktop machine uses a wireless router to capture internet traffic generated by
the development board.
The traffic collected from the router is analyzed to ensure that
benignware and malware execute correctly.  

We use a Samsung Exynos 5250 equipped with a touch screen, and a TI OMAP 5430
development board, and we reboot the boards between each experiment. 
We ran all experiments on the Exynos 5250 because some common apps like NYTimes
and CNN crashed on OMAP 5430 for lack of a WiFi module, but repeated Angry
Birds experiments on the OMAP 5430 to ensure that our results are not an
artifact of a specific device.



\noindent{\bf Performance counter tracing.} We used the ARM DS-5 v5.15
framework and the Streamline profiler as a non-intrusive method for observing
performance counters. 
DS-5 Streamline reads data every millisecond and on every context
switch, so it can ascribe performance events to individual threads.  
However, in DS-5 Streamline extracting per process data can only be done using
its GUI, forcing us to automate this process using the JitBit~\cite{jitbit} UI automation
tool.


\noindent{{\bf Choice of performance counters.} We used hardware
performance counters to 
record five architectural signals: memory loads\textbackslash stores, immediate and indirect control flow instruction counts, 
integer computations, and the total number of executed instructions;
and one micro-architectural signal: the total number of mispredicted branches. 
We collected counter information on a per process basis
as matching programmer-visible threads to Linux-level threads requires
instrumenting the Android middleware (i.e., is non-trivial), and
per-application counters yielded reasonable detection rates. We leave exploring
the optimal set of performance counters for future work.

\noindent{\bf Ensuring correct execution.}  
We ensured that the malicious payload was executed correctly on the board for
each trace. 
Specifically, synthetic malware communicated with a Hercules 3-2-6 TCP
server running on the desktop computer, which recorded a log of all
communication. The synthetic malware itself printed to a console on the desktop
computer (via {\tt adb}) as well as to DS-5 Streamline when running each
malicious payload.

For experiments with off-the-shelf malware, we developed an HTTP server to
support custom (reverse-engineered) duplex protocols for C\&C communication. 
If we allowed malware to communicate to its original server, which was not
under our control, we captured network traffic going through the router. 
We checked the validity of
performance counters readings obtained via DS-5 Streamline with specially
crafted C programs, which we compiled and ran natively on the boards.

\subsection{Constructing and Evaluating HMDs}
\label{sec:train-test}

Using benign and malware traces collected as described above, an HMD analyst
can then train and test a range of HMD algorithms.  For example,
Figure~\ref{fig:ab} shows one of the HMD algorithms we present in a case study
in Section~\ref{sec:results-anomaly}.  The HMD is an anomaly detector and the
figure plots the likelihood that the current trace is going through a known
phase---a low probability thus indicates potential malware (the dark line) while
higher probabilities indicate benignware (light gray lines).
Increasing the payload's intensity lowers the probability even further. By tuning 
the probability at which a time interval is flagged as malicious (or by training
a classifier to learn this), an analyst
can trade-off false positives and true positives.

Importantly, we evaluate true positives and the detection threshold using only
the time windows that contain malware payload execution. We do {\em not} use
time windows where our repackaging code and dispatcher service executes, since
we would like the HMD to be evaluated solely using payloads and not exploits. 
We do not use time windows {\em before} or {\em after} the payload is complete,
because if an HMD raises an alert when the payload is {\em not} executing, the
alert may in reality be a false positive that will get recorded as a true
positive. Prior evaluation methods do not separate out malware payload
intervals and may have this error.  On the other hand, to measure false
positives, we use benign traces only and hence use the entire trace durations
for each experiment. Finally, we use 10-fold cross validation on an appropriate
subset of our data to evaluate HMDs.

\ignore{remove this 10-fold business to results:
We use 5-fold cross-validation with an 80-20 train-test
split~\cite{Liukdd_5_folds,Schatzmann_sigcomm_5_folds} and observe consistent
results across different folds.
We train our HMDs for each baseline application using 80\% of 1-2
hour long benign traces. We do not use any malicious traces to 
train anomaly detectors. 
%
Once trained, we test against the remaining 20\% benign traces to compute the
false positive rate, and test against all malicious traces to compute the
detection threshold and true positive rate for each baseline app. }
%


%
%

\ignore{

One of the goals of our research is to detect repackaged applications, that is
why we validate described above ML methods on such type of applications. For
this purpose, we chose several representative applications (the choice of the
benign apps is discussed in the next section) and injected our malicious code
into them. Repackaging Android applications is very straightforward: first, the
original app is disassembled using apktool utility, second, malicious code is
placed into the proper folders within previously disassembled application,
finaly, all files can be assembled using the same apktool and signed using
jarsigner tool, and a new apk archive is ready to be deployed. The simplicity
of this process partially explains why there are so many repackaged Android
applications on 3rd party markets.

\subsection{Malware design}

We considered a few strategies of organizing the malicious code itself. For
example, each basic action, such as stealing files, DDoS, and device IDs, could be run sequentially or in parallel; they can stay within a single
Android activity or service, or for each of them a separate activity/service
can be launched. Based on the analysis of real malware samples, we decided to
execute each basic malicious activity within a separate asynchronus service. We
also provided an opportunity to launch several parallel threads in order to
maximize the throughput of time-consuming actions. For example, network-related
malicious activities (e.g., click fraud) may block on network accesses. All
malicious functionality is implemented within one program. As an input, it
takes a configuration file that defines what actions should be carried out and
parameters of command \& control server (C\&C). C\&C server is used to receive
data transmitted by malware, thus simulating network communication of an app
with an Internet server. It can also communicate with our repackaged app, but
we did not use this functionality in our experiments because it adds extra
complexity, but it does not affect the results. As we noticed when analyzing
Android apps, most of them are obfuscated using standard Proguard tool. We also
applied Proguard to the source code of our malicious program.

\subsection{Experimental setup}

\begin{figure}[tbp]
\centering
\includegraphics[width=0.45\textwidth]{figs/environment.pdf}\par
\caption{Experimental setup}
\label{fig:exp_setup}
\end{figure}

The major parts of our experimental setup are shown in Figure ~\ref{fig:exp_setup}. 
One of the challenges was to collect data from performance counters. We were looking for a nonintrusive way of observing performance counters and matching this information with OS-level activities (i.e., contex switches, thread blockng events, etc.). ARM DS-5 framework equipped with Streamline profiler v5.15 seemed to be a good fit for our problem because it is officially supported by ARM and can work with multiple ARM-based development boards. In our experiments, we used two development boards: Samsung Exynos 5250 equipped with Cortex-A15 processor, and TI OMAP5430. On both boards, we installed Android JellyBean v 4.1.1 (Linux kernel 3.0.31).

All experiments were conducted on either OMAP 5 development board or on ARndale Exynos board, but we mostly used the latter board because the former one has issues with running some Android apps. The boards were connected to the local computer, and performance counters were accessed using local installation of DS-5 Streamline ~\ref{} software. Streamline consists of two parts: a driver that is installed on a board, and a UI component that communicates with the driver and helps to configure it. Streamline collects raw data and stores it on a local computer. Processing data and extracting per-process performance counters requires lots of RAM and computational resources. In order to speed up this stage, we transferred data to the server. When running Android malware, the same sequence of user actions was replayed for all malware samples injected into an app. We used AndroidReran ~\ref{} application to record and replay user actions. Both AndroidReran and DS-5 Streamline were coordinated by the local computer. One of the major shortcomings of DS-5 Streamline is that it requires us to use a GUI interface to extract per-process counters. As a workaround, we used JitBit ~\ref{} UI automation tool. And, finally, when per-process performance counters were extracted, they were used as an input data for our ML algorithms.

\begin{table}[t] 
\caption{Performance counters} 
\centering
\begin{tabular}{l l l}
\hline\hline
Counter name & ARM index \\ 
\hline 
Branch: Mispredicted   & 0x10 \\
Load/Store inst. & 0x72 \\
Integer inst. & 0x73 \\
Indirect branch inst. & 0x7a  \\
Imm. branch inst. & 0x78  \\
Executed inst. & 0x08 \\
Cycles & - \\ 
\hline 
\end{tabular} 
\label{table:perf_counters} 
\end{table} 

\subsection{Performance counters}
In our experiments, we inclined toward using the local counters -- i.e., the counters that can characterize individual processes. However, we used one global counter that counts the number of mispredicted branches. The set of counters used for the remaining experiments and their ARM indices are shown in fig. ~\ref{fig:perf_counters_choice}. Preliminary experiments showed that the chosen set of counters characterizes Android applications well and is applicable for building dynamic signatures. We believe that collecting counter information on the per-process basis is the most efficient way of constructing computational signatures. Per-thread data is too fine grained and does not permit us to see the global picture. Streamline reads data every millisecond or on every context switch, which is why it can ascribe performance events to individual processes and even threads.
}


\section{Case Studies using EMMA}
\label{sec:results}
We show how malware analysts can use \name through
three case studies. 
{\bf (1)} We use malware payload sizes in Section~\ref{sec:malware-gen}
to tune the machine learning features (100ms v. sub-ms in prior work) 
for an anomaly detector HMD. 
Our HMD out-performs prior work designed to detect short-lived exploits 
by 24.7\% on the area under curve (AUC) metric (Section~\ref{sec:results-anomaly}).
{\bf (2)} \name's
taxonomy of malware in Section~\ref{sec:malware-gen} can be used to train a supervised learning based
HMD eficiently. This `balanced' HMD outperforms alternative HMDs -- that are trained
on subsets of malware behaviors -- when tested on new variants of the behaviors.
{\bf (3)} Surprisingly, we show that our anomaly-based HMD can detect malware
that uses obfuscation to evade the best (deployed as well as in research) static analyses. 
Hence, HMDs and static analyses are complementary and can drive malware payloads towards inefficient 
implementations.

\subsection{Anomaly detector using \name's taxonomy}
\label{sec:results-anomaly}


\begin{figure}[t]
  \begin{minipage}[t]{0.5\linewidth}
    \centering
    \includegraphics[width=\textwidth]{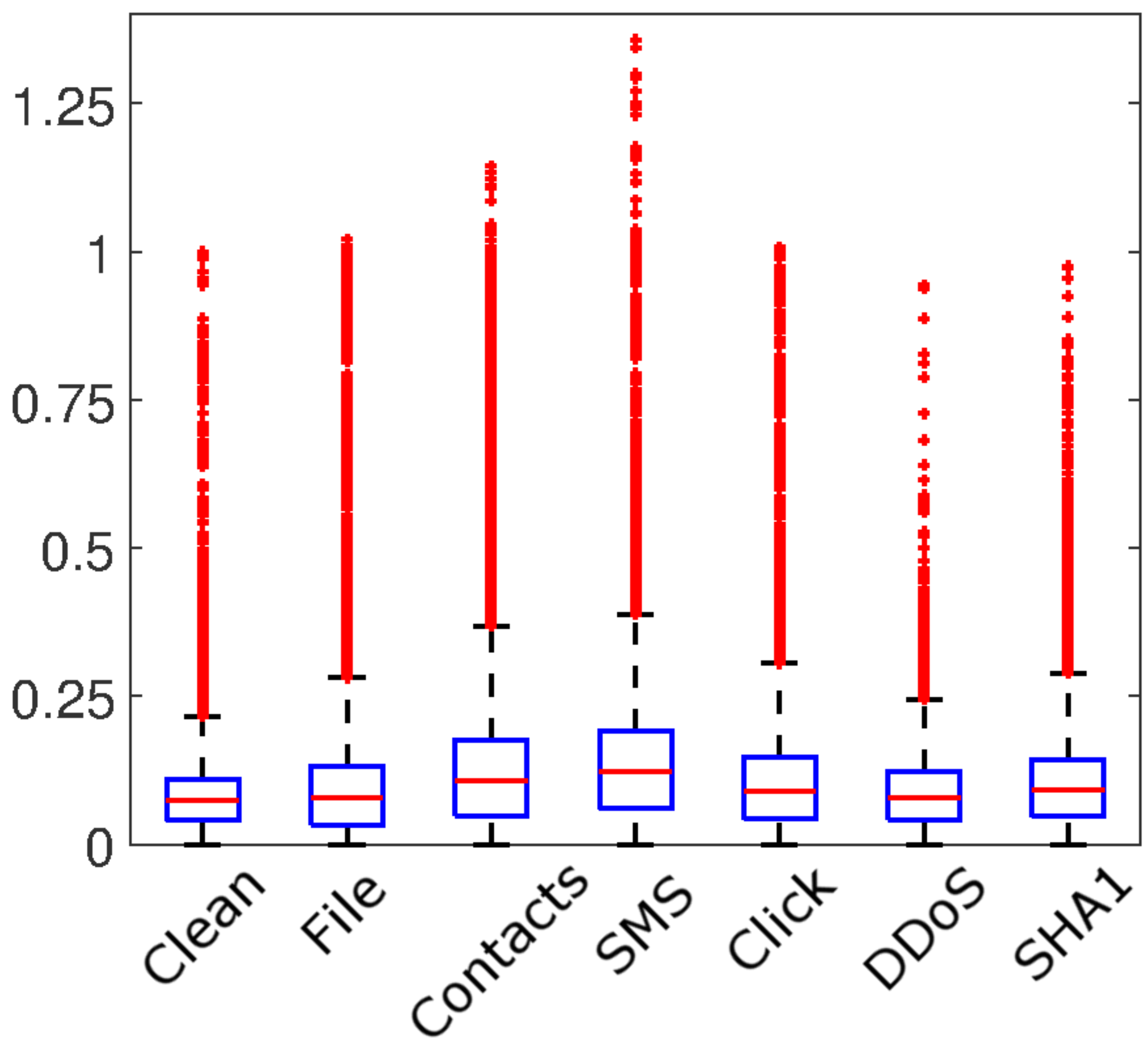}
\label{fig:sign-roc}
  Before
  \end{minipage}
  \begin{minipage}[t]{0.45\linewidth}
    \centering
    \includegraphics[width=\textwidth]{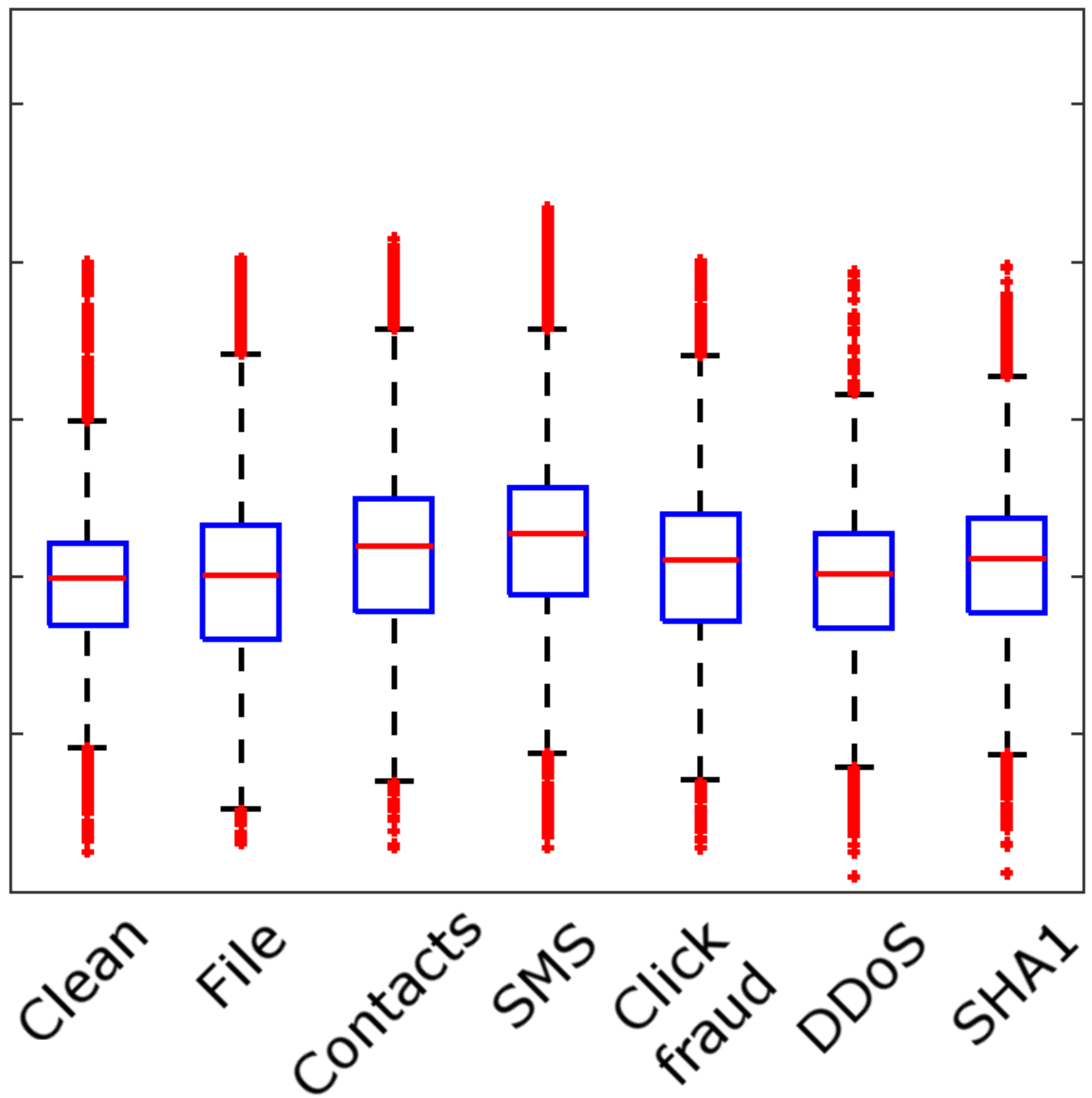}
\label{fig:sign-auc}
  After
  \end{minipage}  
  \caption{ Distribution of load/store events in Angry Birds 
before and after power transform. 
Power transform does not make malware {\em payloads on Android} more discernible from benign behavior,
whereas Tang et al.~\cite{anomaly-detector-columbia-raid} show that it separates {\em exploits} from
benign apps in Windows.
}  
\label{fig:power-trans}
\end{figure}

\begin{figure}[tbp]
   \centering
   \includegraphics[width=0.45\textwidth]{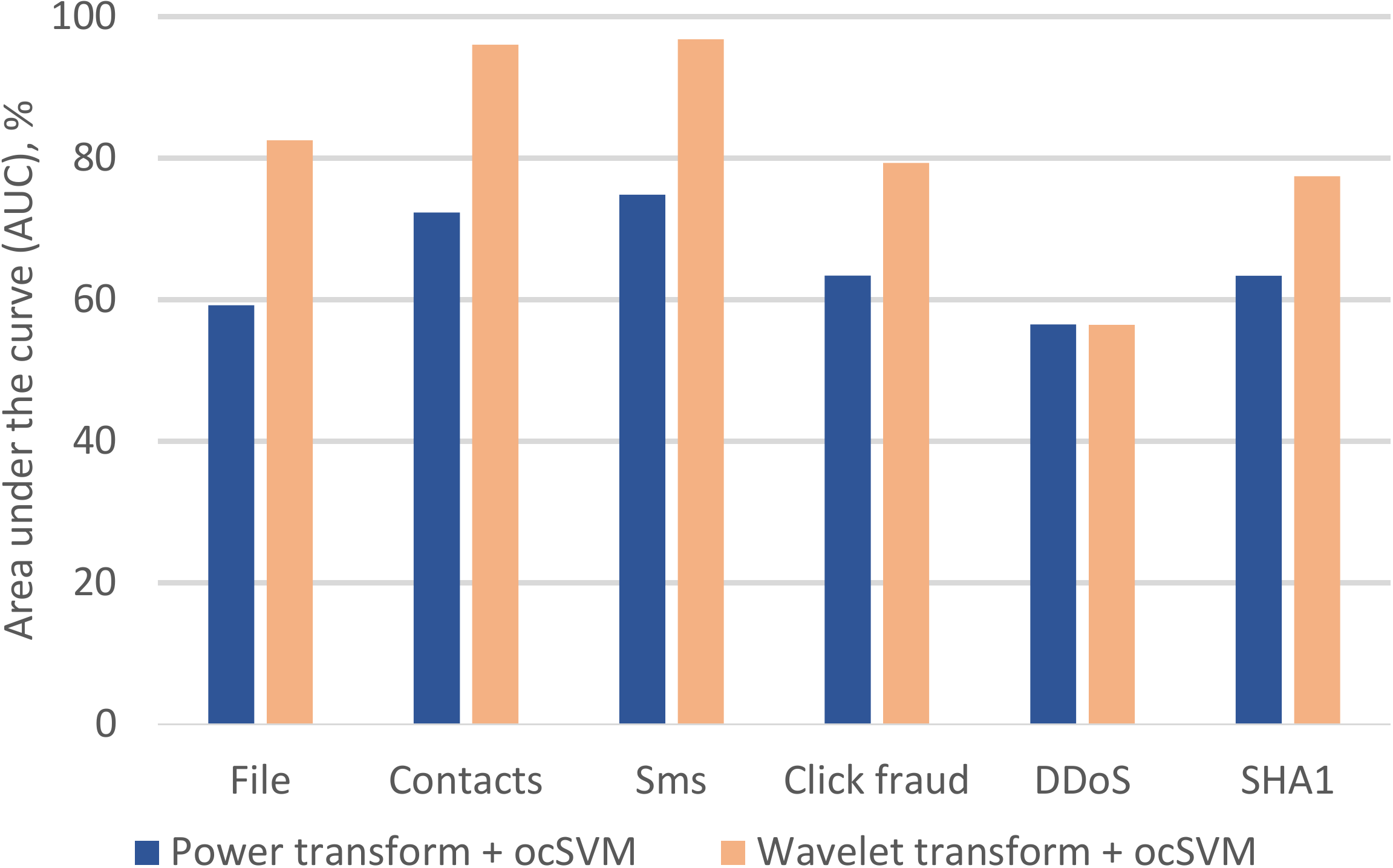}\par
\caption{
Comparison of power transform + ocSVM (prior work) and Discrete Wavelet Transform + ocSVM (this work). Our detector has 24.7\% better area under curve metric (AUC)
than prior work.}
\label{fig:power-v-dwt}
\end{figure}

We begin by quantifying why 
prior work designed to detect exploits may not yield the best HMDs
to detect long-lived payloads. 

\noindent{\bf Exploit-based ML features do not expose payloads
(Figure~\ref{fig:power-trans}).} Tang et al.~\cite{anomaly-detector-columbia-raid} present an HMD
specifically designed to detect the multi-stage exploits that characterize
Windows malware. The HMD samples performance counters every 512k cycles, and
uses a power transform on performance counter data to \ignore{fit it into a
Gaussian distribution and thus}separate benign and malicious time
intervals.  Then, a one-class SVM (ocSVM) is trained on short-lived features --
i.e., on each sample as a non-temporal model and using 4 consecutive samples to
train a temporal model -- to label anomalous time intervals as malicious. 

We find that power transform does {\em not} have the same effect on mobile
malware payloads---payloads look very similar to benignware traces even after a
power transform.  For example, Figure~\ref{fig:power-trans} shows the
distribution of load-store instruction count per time interval for benign Angry
Birds (labeled `Clean'), compared to time intervals in Angry Birds infected
with different malware payloads (e.g., file stealer, click fraud, DDoS,
etc)---before and after a power transform.  The distributions are shown as a
box-and-whiskers plot, where the box edges are $25^{th}$ and $75^{th}$
percentiles, the central mark is the median, the whiskers extend to the most
extreme data points not considered outliers, while the outliers are plotted
individually in red.  Data in both plots have been normalized to the range of
benign Angry Birds' values.  We use the standard Box-Cox power transformation
to turn performance counter traces into an approximately normal distribution.
Since the distributions of malware and benignware in
Figure~\ref{fig:power-trans} overlap significantly, training an ocSVM on this
dataset will yield a poor HMD as we show next.

\ignore{discuss each box: mean, std deviation, percentile, etc.}
\ignore{Mikhail: what is the standard process for computing the ideal lambda?
Discuss here.}


\noindent{\bf Payload-centric ML features.} We designed a new HMD
whose features reflect our findings about mobile malware payload sizes in
Figure~\ref{fig:synthetic-malware}.  
Specifically, we attempt to capture program effects at the scale of 100ms
intervals, i.e., closer to the time required for atomic actions like stealing
information or networking activity.

We then extract features from each 100ms long time interval using Discrete
Wavelet Transform (DWT) and use the wavelet coefficients as a feature vector
for the time interval.  The wavelet transform can provide both accurate
frequency information at low frequencies and time information at high
frequencies, which are important for modeling the execution behavior of the
applications.
We use a three-level DWT with an order 3 Daubechies wavelet function (db3) to
decompose a time interval. We also used the Haar wavelet function, but did not
observe much difference in the detection results.


Finally, we use multiple feature vectors to construct two models:
(a) a bag-of-words algorithm followed by a ocSVM, and (b) a probabilistic
Markov model. Both these models are simple to train and compute at run-time,
and hence serve as good local detectors 
(and a good baseline for more complex models such as neural nets that are
harder to train). 




\subsubsection{Bag-of-words Anomaly Detector}
\label{sec:bow}

The bag-of-words model treats 100ms time intervals as words and a
Time-to-Detection (TTD) window as a document.  We experimented with a range of
words and TTDs, finding a codebook of 1000 words and TTD = 1.5 seconds to yield
good results. The bag of words algorithm maps each TTD window into a 1000-entry
histogram, and trains a one-class SVM on benign histograms.  We parameterize
the one-class SVM so that it has $\sim$20\% percent false positives. 


\noindent{\bf Comparison with power transform | ocSVM HMD.}
Figure~\ref{fig:power-v-dwt} compares our bag-of-words based
ocSVM with one that uses a power transform using the area under
ROC curve (AUC) metric. Note that AUC is a relative metric to 
compare classifiers, whereas the operating range measures an HMDs'
robustness to atomic-action-sized mutations in malware.
The bag-of-words model outperforms prior work 
for each category of malware behavior
and by an average of 24.7\% higher AUC across all malware.  


\noindent{\bf Operating range of DWT | bag-of-words | ocSVM.}
Figure~\ref{fig:synthetic-SVM-heat-map} shows the operating range for 
the bag-of-words model.
Each cell in the matrix corresponds to a malware payload
action (y-axis) and benign app (x-axis) pair. The malware payloads are grouped by
category and within each category, increase in size from top to bottom
and in delay from right to left.  
These experiments use parameters from 
Figure~\ref{fig:baseline-apps}.
The intensity of the color -- from light green to dark red --
corresponds to the detection rate, which is computed as the number of raised
alarms versus the total number of alarms that could be raised. 


Figure~\ref{fig:synthetic-SVM-heat-map} shows that the bag-of-words model
achieves, at $\sim$20\% false positive rate: 1) surprisingly high true positive
rate for dynamic, compute intensive apps such as  Angry Birds (99.9\%), CNN
(84\%), Zombie WorldWar (93\%), and Google Translate (92.4\%); and 2)
$\sim$80\% true positive rate for both Amazon and Sana. 

\begin{figure}[tbp]
   \centering
   \includegraphics[width=0.45\textwidth]{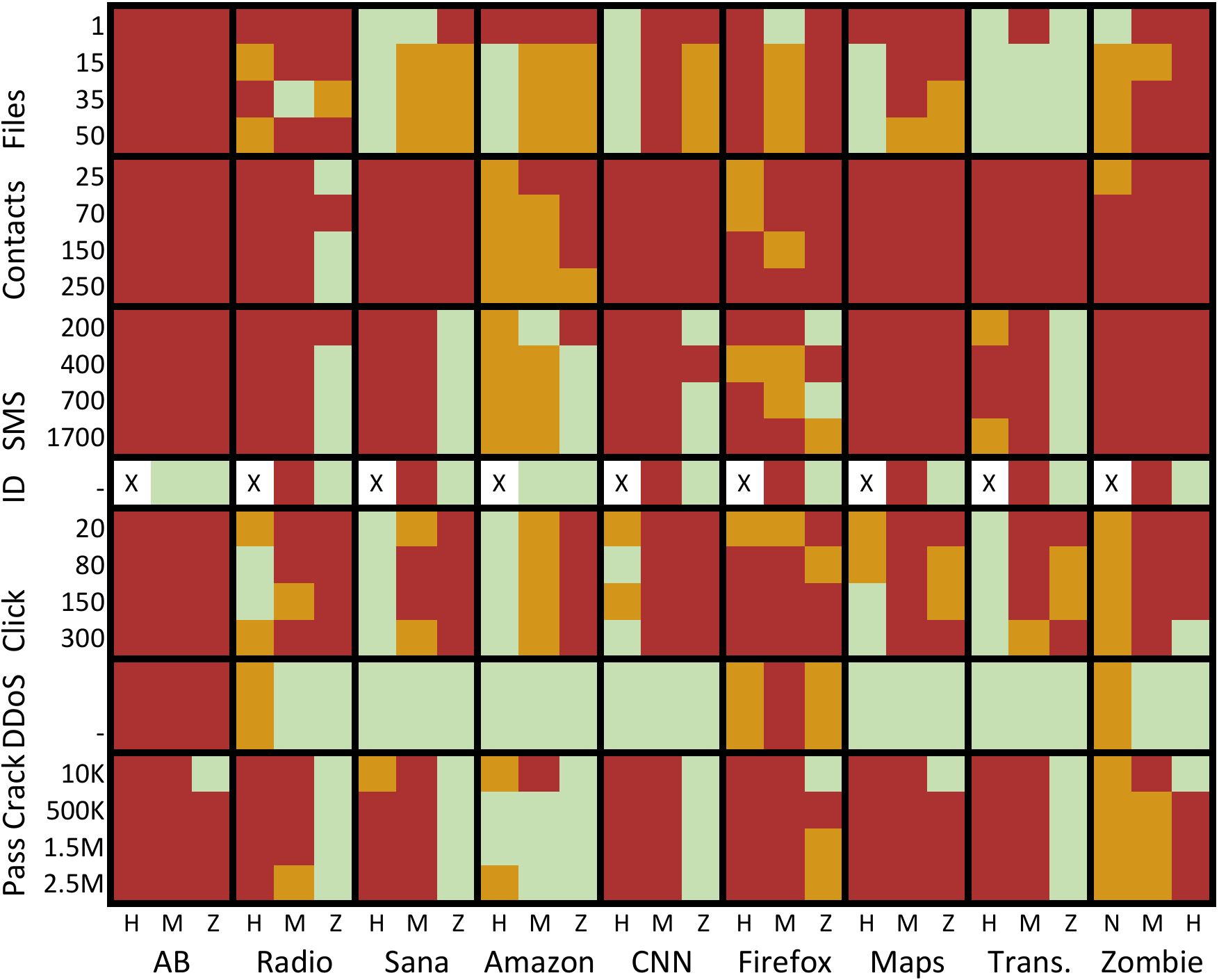}\par
\caption{
The operating range of Bag-of-words HMD.  
In each rectangle, the size of malicious payload grows from
the top to the bottom, and the amount of delay decreases
from left to right (H=High, M=Medium, Z=Zero delay). If color goes from light to dark within a rectangle,
then the detection threshold (i.e., the lower end of the operating range) lies inside the rectangle. 
}
\label{fig:synthetic-SVM-heat-map}
\end{figure}

\subsubsection{Markov-model based Anomaly Detector}


We present an alternative HMD to show that HMD models should be chosen specific
to each application, and that there is an opportunity to apply ensemble methods
to boost detection rates.

Our first-order Markov model based HMD assumes that the normal execution of an 
application (approximately) goes through a (limited) number of states 
(program phases), and 
the current state depends only on the previous state. The goal is to detect 
malware if its performance counter trace creates 
a sequence of rare state transitions (as shown in Figure~\ref{fig:ab}). 

The HMD uses DWT to extract features as in the bag-of-words model, but maps
them to a smaller number of words (i.e., states in the Markov model) using
k-means clustering.  We use the Bayesian Information Criterion (BIC)
score~\cite{Pelleg-xmeans} to find that 10 to 20 states is a good number
across all benign apps. 
Using observed state transitions derived from 
training data, we empirically estimate the 
transition matrix and initial probability distribution (through
Maximum Likelihood Estimation).  
For detection, the Markov model HMD tracks the joint probability of
a sequence of states over time and if malware computations create anomalous
hardware signals (i.e. this probability is below a threshold for 5 states in our model),
the HMD raises an alert.

\begin{figure}[tbp]
   \centering
   \includegraphics[width=0.45\textwidth]{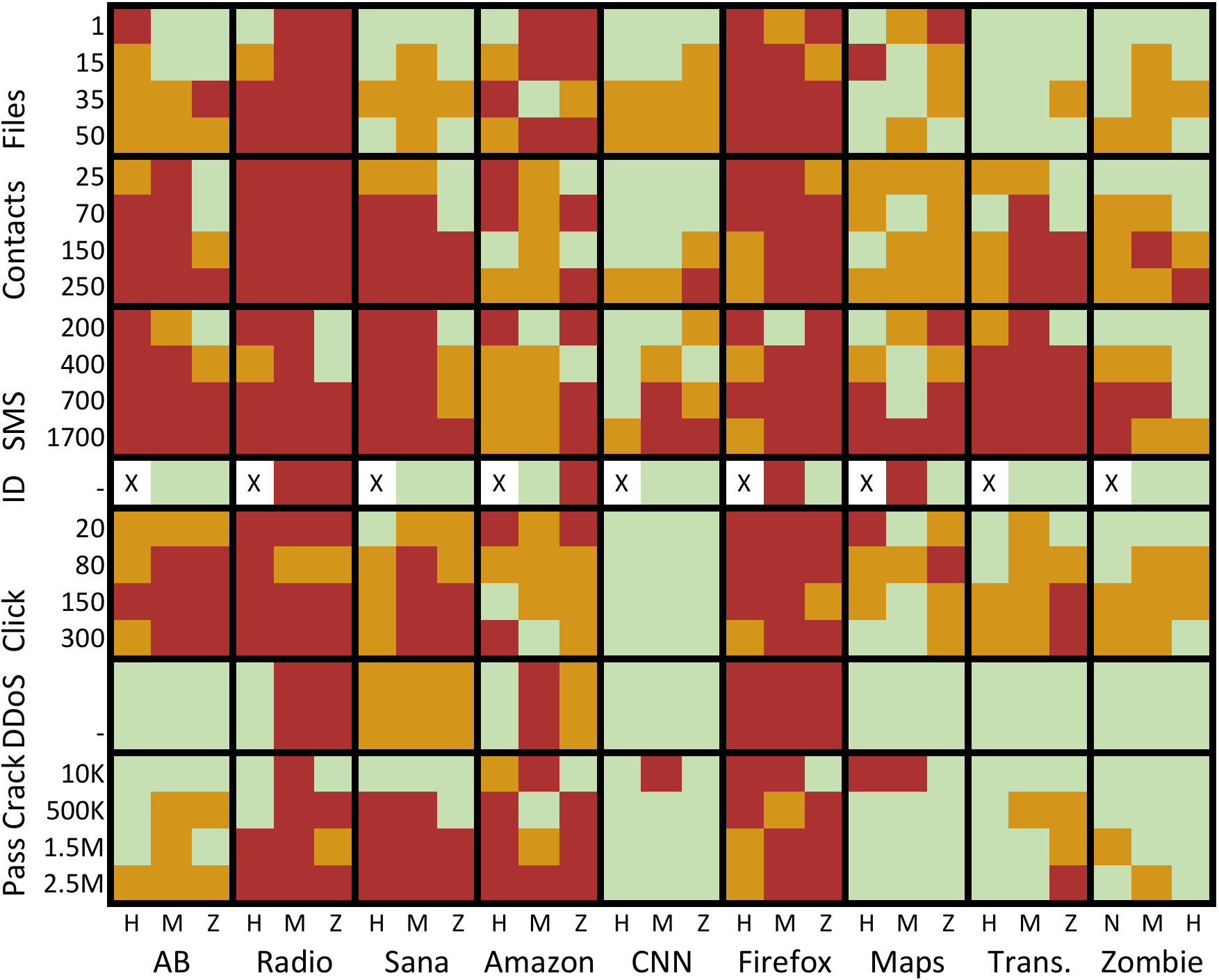}\par
\caption{The operating range of Markov model HMD. 
Interestingly, the Markov model performs worse than the simpler
bag-of-words model for compute intensive and dynamic apps (e.g., Angry Birds, CNN, and Zombie
WorldWar). 
}
\label{fig:synthetic-MM-heat-map}
\end{figure}

\noindent{\bf Operating range of DWT | Markov model HMD.}
Figure~\ref{fig:synthetic-MM-heat-map} shows the results-matrix for the Markov
model based detector. All the results are shown for a false positive rate of
20-25\%.

Increasing the size of each payload action makes malware more detectable --
this can be seen as the colors being more intense towards the bottom part of
most rectangles.  Increasing the delay between two malicious actions does not
have a similarly predictable effect -- SMS stealers in Angry Birds is a rare
pair where detection rate increases with delay. This is interesting since
intuitively, adding delays between payload actions should decrease the chances
of being detected. However, these experiments indicate that for most
malware-benign pairs, detection depends on how each payload action interferes
with benign computation rather than delays between the payload actions.  

The most important take-away from Figure~\ref{fig:synthetic-MM-heat-map} is
that for most malware-benignware pairs, the detectability changes from light
green to dark red as we go from top to bottom in the rectangle -- this shows
that our malware parameters in Figure~\ref{fig:synthetic-MM-heat-map} are close
to the detection threshold, i.e. the lower end of the HMD's operating range for
the current false positive rate.  There are a few exceptions as well, such as
click fraud, DDoS, and password crackers hiding in CNN; and DDoS in Angry
Birds, Maps, Translate, and Zombie World Wards. For these cases, the payload
intensity has to be increased further to find their detection threshold.

\noindent{\bf Markov model HMD space and time overheads.} Markov models representing
the behavior of the benign apps vary from 1.2KB to 6.7 KB, with an average size
of 3.2KB -- they are thus cheap to store on devices and transfer over cellular
networks.  Its time to detection ranges between 1.2 seconds to 4.4 seconds and
about 2.5 seconds on average.  This means that the system can detect suspicious
activities at the very beginning, considering that exfiltrating even one photo
takes 2.86 sec on average.

\subsubsection{HMDs should be app-specific}

Interestingly, the Markov model works significantly better than bag-of-words 
for TuneIn Radio -- with a 10\% FP: 90\% TP rate compared to 38\%FP: 90\% TP
rate respectively -- but performs significantly worse on apps like Angry Birds.
In summary, a deployed HMD will benefit from choosing the models that work best
for each application, but due to their different TP:FP 
operating points, will also benefit from
using boosting algorithms in machine learning~\cite{schapire2012boosting}.

\subsection{\name improves accuracy of supervised 2-class HMDs}
\label{sec:results-signature}

\begin{figure*}[t]
  \begin{minipage}[t]{0.45\linewidth}
    \centering
    \includegraphics[width=\textwidth]{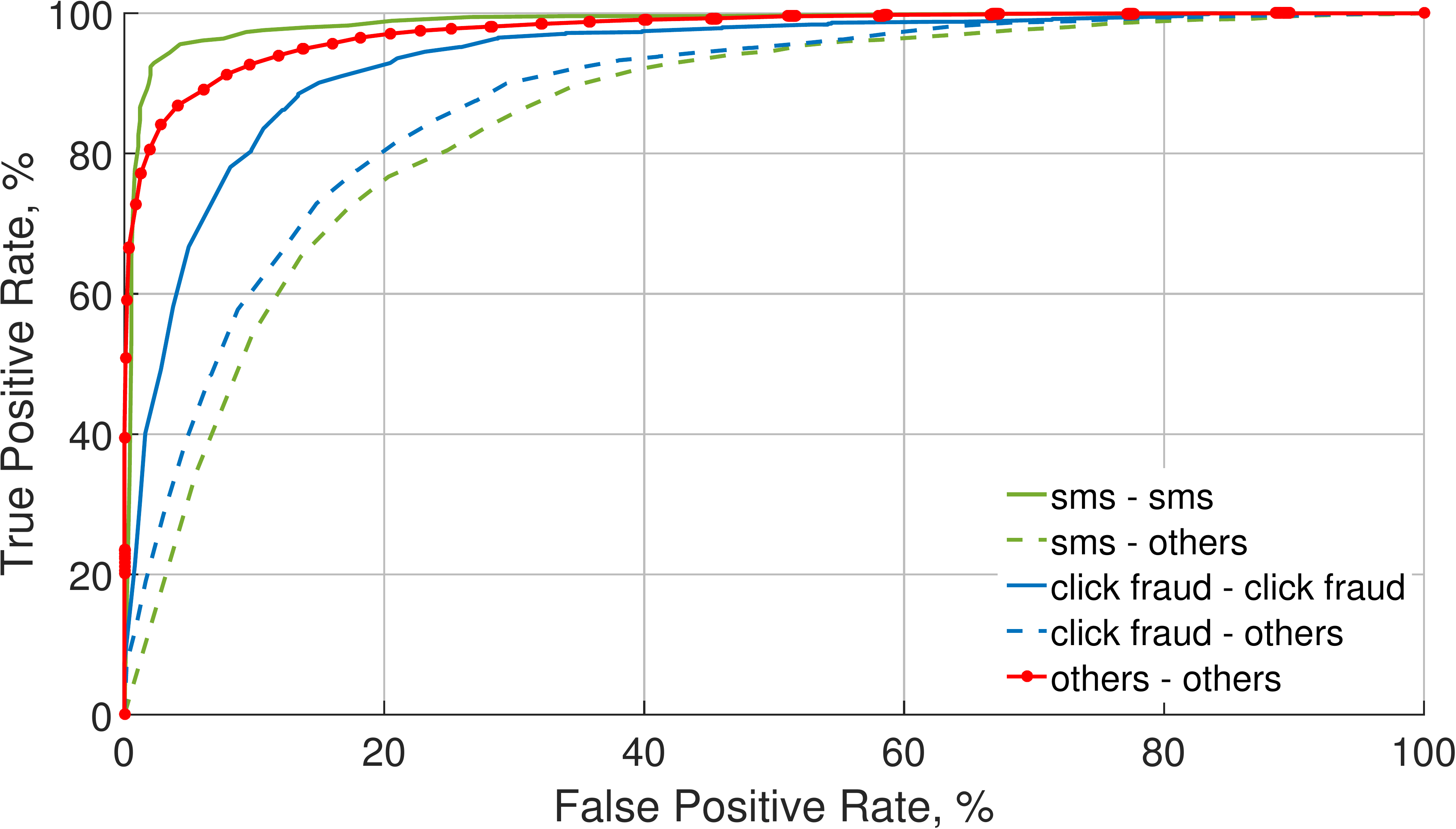}
\label{fig:sign-roc}
  \end{minipage}
  \begin{minipage}[t]{0.45\linewidth}
    \centering
    \includegraphics[width=\textwidth]{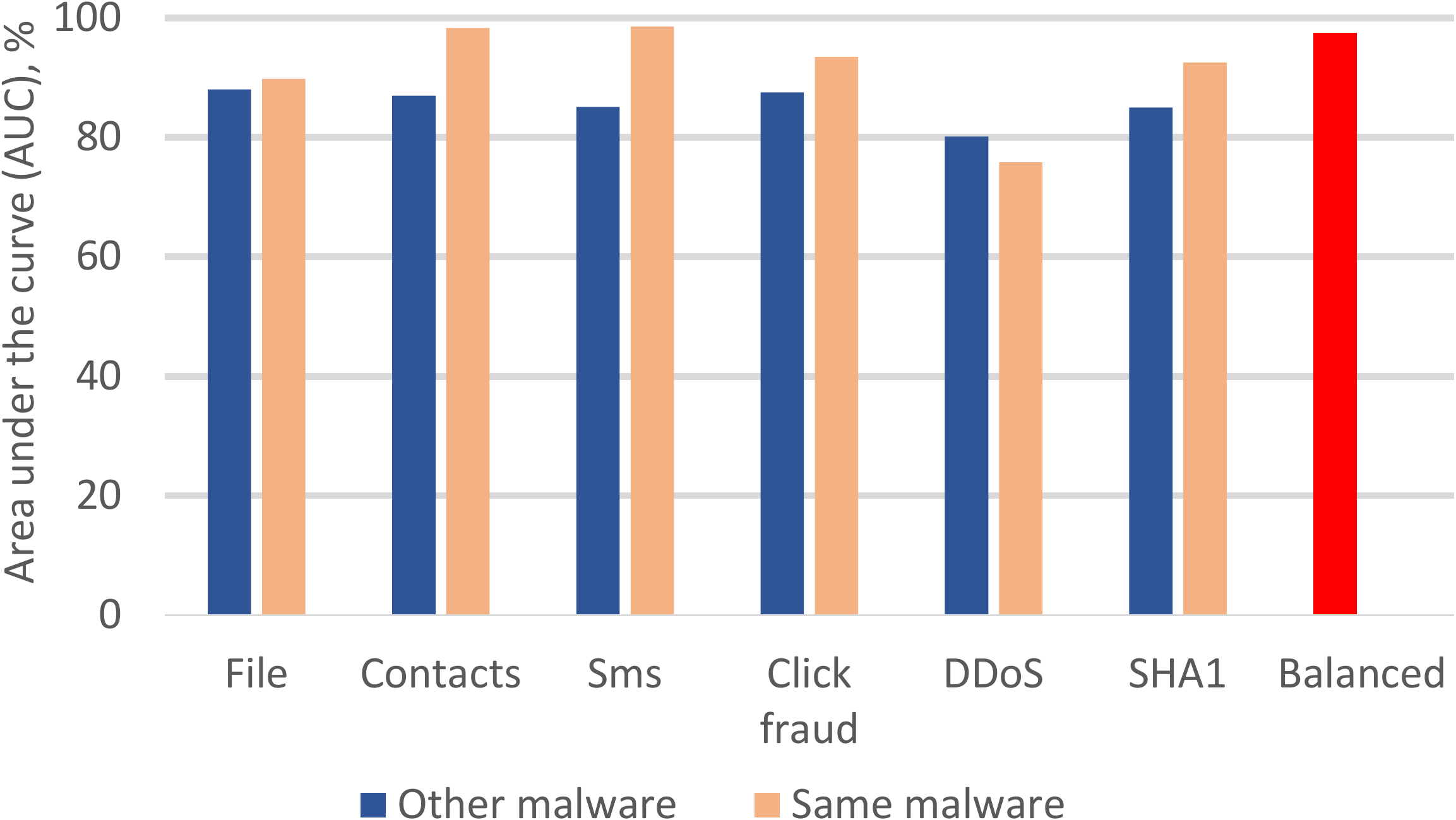}
\label{fig:sign-auc}
  \end{minipage}  
  \caption{Training supervised learning HMD on a balanced set of malware behaviors yields best results.}  
\label{fig:sign-failure}
\end{figure*}

\begin{figure}[tbp]
   \centering
   \includegraphics[width=0.45\textwidth]{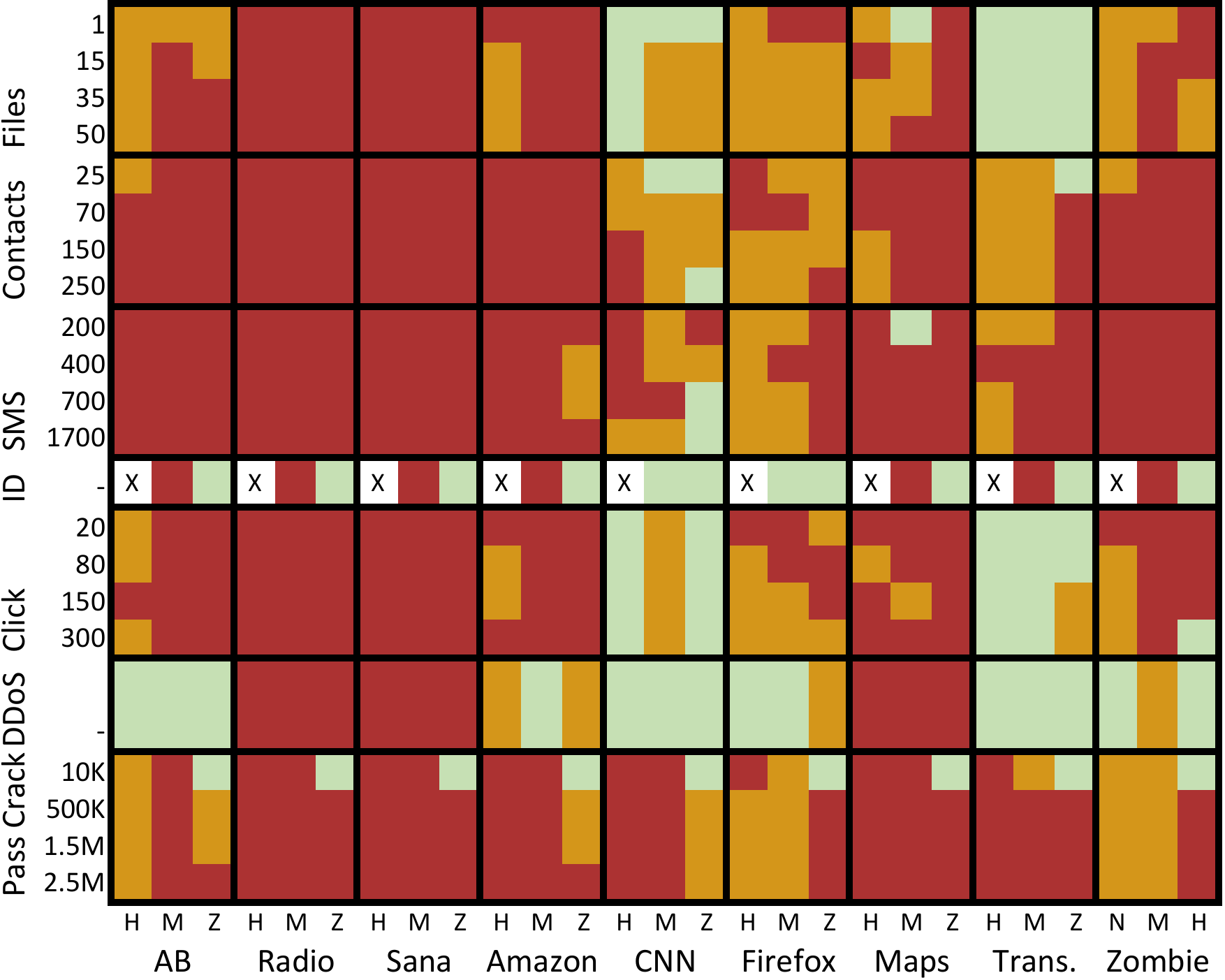}\par
\caption{Operating range of 2-class Random Forest HMD: more effective than
anomaly detectors when trained on a balanced dataset of all malware behaviors.}
\label{fig:sign-heatmap}
\end{figure}

\ignore{
Seeing any variant of a malware behavior improves a signature-based HMD's 
detection rate when compared to an HMD that hasn't been trained on 
the overall behavior. This shows the utility of our behavioral taxonomy.

A representative training set that contains all behaviors -- some variants
of each behavior which are not the same as the variants in the testing set -- 
improves the detection rate.

Experimental setup: HMD = features, model, time intervals, etc.

Plots in Fig 14.

Fig 15: compare signature to anomaly operating range.

}

\name can significantly improve the performance of supervised learning based HMDs;
specifically,
by training the HMDs on a `balanced' training data set that contains 
malware with each high-level behavior (Figure~\ref{fig:sign-failure}).
Note that supervised learning techniques can be trained to recognize specific families
of malware~\cite{simha-isca} (i.e. a multi-class model) 
or to coalesce all feature vectors into one label (i.e., a 2-class model)---we evaluate
both categories in Figure~\ref{fig:sign-failure}. 

In all the following experiments the number of training samples is fixed to exclude
bias from training sets of different size. Each training set is balanced, i.e.
contains equal number of benign and malicious samples. 
The results are computed using 10-fold cross validation.



We experimented with several supervised learning algorithms -- e.g., decision tree,
2-class SVM, k-Nearest Neighbor, Boosted decision trees, and Random Forest (RF)-- and 
present the results for RF classifier because it
demonstrated the best performance on our data set. 
In Figure~\ref{fig:sign-failure}, we present results using ROC curves (left) and AUC metric (right)
to compare relative performance of RF under different
training and testing data sets.
Solid lines in the ROC plot correspond to testing on the same malware type that is used for
training, while the dashed lines show RF's performance on other
malware types.

The common trend that we observed across all nine apps and all
malware types is that the RF classifier has
significantly better performance when testing on the same malware types (solid
lines are higher than the dashed ones). 
The only exception is when the RF HMD is trained on DDoS malware, it surprisingly achieves better
performance on other malware behaviors than on the in-class malware
behaviors. 



Further, we trained a classifier on a balanced set of malicious data that
included all malware behaviors in \name. The solid line with dots (in the ROC
plot) and the column on the far right (in the AUC bar graph) in
Figure~\ref{fig:sign-failure} show that showing some variants of each behavior
enables the RF to achieve a higher detection rate (on even new variants) than
both prior work as well as one-class SVMs. 
The RF HMD can, for example, detect close to 85\% of the malware with only 5\%
false positives compared to our anomaly detectors' similar true positives for 
$\sim$20\% false positive rates. Finally, the 
RF HMD trained on a balanced data set 
yields 97.5\% AUC 
whereas RF HMDs trained on per-behavior inputs yield AUCs of 91\% and 85\% 
when tested against the same or new malware behaviors respectively (averaged across
all behaviors).


\noindent{\bf Operating range of Random Forest HMD.}
Figure~\ref{fig:sign-heatmap} shows the detection results matrix for the RF HMD
across the entire malware payload (Y-axis) and benignware (X-axis) categories
for a fixed false positive rate of 5\%. The key results are that RF detects
most payloads except for detecting click fraud and DDoS attacks in CNN,
Firefox, and Google Translate. It is likely that DDoS attacks -- which involve
a sequence of infrequent HTTP requests -- look very similar to benign apps and
are not well suited to be detected using HMDs. Indeed, all three HMDs --
bag-of-words, Markov model, and RF -- do a poor job of detecting DDoS attacks
in most apps. On the other hand, RF consistently detects information stealers
and compute malware (password cracker) across most apps. For apps with regular
behavior (Radio) or sparse user-driven behavior (Sana), RF can detect all but the 
smallest of malware payloads.

In summary, \name helps an analyst develop a robust HMD---first by dissecting
existing malware to identify orthogonal behaviors, and then by training the HMD on
a representative set of malicious behaviors. In the end, using the operating range,
\name informs the analyst of the type of behaviors the HMD is well/poorly suited
at detecting.

\subsection{Composition with Static Analyses}
\label{sec:reflection}

\begin{figure}[tbp]
\input{malware}
\caption{Code shows Java reflection and string encryption
in {\tt Obad} malware that foils static analysis tools.}
\label{fig:reflection-code}
\end{figure}
\mohit{Future work: Bag of words model on the Obfuscated traces?}

Reflection is a powerful method for writing malware that 
evades static program analysis tools used in App Stores
today~\cite{evade-bouncer}. Interestingly,
we show that malware that uses
reflection to obfuscate its static program paths
in turn worsens its dynamic hardware signals, and 
improves HMDs' detection rates.
 
Java methods invoked via reflection are resolved at runtime, making it hard for
static code analysis to understand the program's semantics.
At the same time, reflection alone is not  sufficient -- all strings in the
code must also be encrypted, otherwise the invoked method or a set of possible
methods might be resolved statically.  \ignore{Code snippet of
reflection+encryption and its effect on Flowdroid like static analyses.} 
%
\begin{figure}[tbp]
   \centering
   \includegraphics[width=0.45\textwidth]{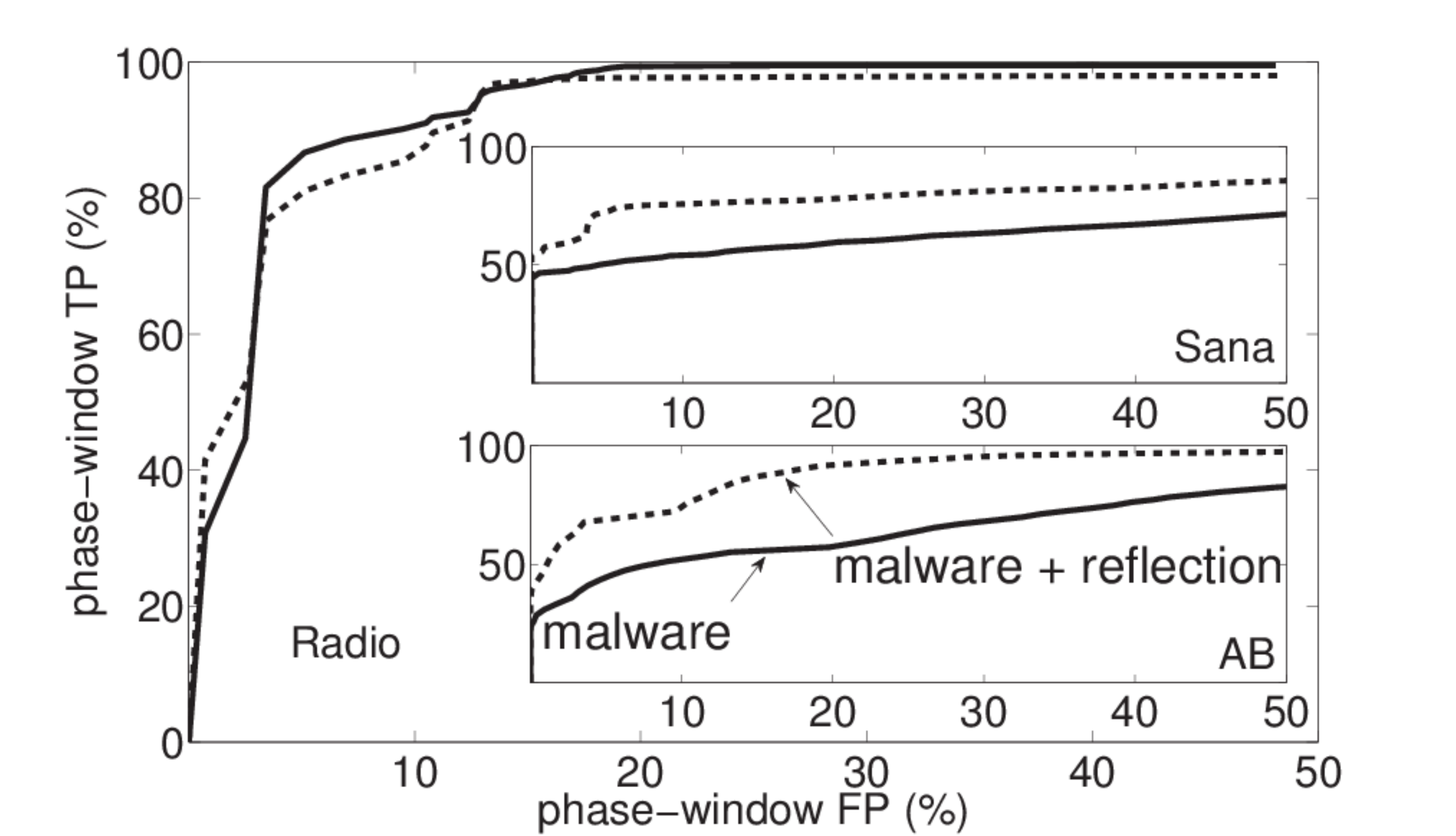}\par
\caption{(Markov model) Effect of obfuscation and encryption on detection rate: interestingly, malware becomes more distinct compared to baseline
benign app.}
\label{fig:reflection}
\end{figure}

To illustrate an actual malicious use of Java reflection and encryption, we
show a code snippet (Figure~\ref{fig:reflection}) from {\tt Obad}
malware~\cite{obad}. The code decrypts class and method names (lines 6
and 10) by calling the method {\tt oCIlCll()}.  As a result, static 
analyses~\cite{flowdroid,chex} either do not model reflection
or conservatively over-approximate the set of
instantiated classes for {\tt method\_name} (line 10)
and target methods for the {\tt invoke} function (line 14). 
Due to control-flow edges that may never be traversed, static data-flow analysis 
becomes overly conservative, and
static analyses end up with high false positive rates (or more commonly, with malware
that goes undetected).


We augmented our synthetic malware with reflection and encryption similar to
{\tt Obad}'s implementation.  Static analysis of our malware does not reveal
any API methods that might raise alarms---we tested this using the Virustotal
online service which ran 38 antiviruses on our binary without raising any
warnings. 

Figure~\ref{fig:reflection} shows results of using the Markov model HMD on the 66 synthetic
malware samples from Figure~\ref{fig:synthetic-malware} 
augmented with reflection and encryption, and embedded into each of 
AngryBirds, Sana, and TuneInRadio.
%
We see that in Angry Birds and Sana the detection rate of the malware that uses
both reflection and encryption is significantly higher because reflection and
encryption are computationally intensive and disturb the trace of the benign
parent app (i.e., more than the same malware without reflection and
encryption).  We do not see the same trend for TuneInRadio because its
detection rate was already quite high, so the additional impact of reflection
on TuneIn Radio stays within the noise margin. 
We conclude that HMDs complement current static analyses and can
potentially reduce the pressure on
computationally intensive dynamic analyses with a larger trusted code
base~\cite{taintdroid}.

\vspace{-0.1in}
\section{Conclusions}
\label{sec:conclusion}


\name is particularly relevant to computer architects
since hardware behaviors in payloads are easy to obfuscate, whereas
malware semantics are forced to be more consistent at the software level.  
This paper
focuses on diversifying hardware signals and evaluating hardware 
detectors---future work will look into 
applying \name's principles to software detectors.

Our results show that 
HMDs, just like system call based behavioral detectors~\cite{mihai-issta}, 
have false positives rates that preclude solo
deployment. However, HMDs form a trustworthy local detector that can be
isolated from kernel compromises or user errors---our results show that
HMDs separate out
true and false positives well enough for a global detector to be
triggered on-demand and apply distributed algorithms to boost
the global malware detection rate~\cite{dash06}. 

HMDs complement prior work in static and dynamic analysis of mobile malware.
Static analysis research includes analyzing apps' permissions to detect overprivileged
apps~\cite{android-demystified}, detecting malware repackaged in benign apps
through pairwise comparison of
binaries~\cite{DNADroid,DroidMOSS,Juxtapp,PiggyApp}, or using
type-systems~\cite{ernst-taint-type} and program dependency
graphs~\cite{apposcopy} to detect malware.  
Dynamic analyses observe an app's run-time data to enforce access control or
information flow
policies~\cite{DroidScope,XuSA2012,HornyackHJSW2011,taintdroid}.
Aurasium~\cite{XuSA2012}
repackages existing apps by attaching user-level sandboxing and
policy-enforcement code.  

 
Our future work will include composing HMDs with system call and compiler-runtime
based malware detectors.
Our approach of identifying {\em why} a detector succeeds and fails,
instead of black-box experiments with malware binaries, is crucial towards this
goal. Indeed, prior work has pointed out the pitfalls of using machine learning
in a black-box manner for network-based intrusion detection systems~\cite{paxson1999}---we
provide a framework to conduct evaluate behavioral detectors in a principled manner.
Purely machine learning-based ensemble methods to compose weak detectors
into one robust detector are also a rich vein of research to draw from~\cite{ponomarev-ensemble-learning}.

Finally, computer architects are exploring
new hardware signals and accelerators to improve security in general and
malware detectors in particular---our
work lays a solid methodological foundation for future research into
HMDs for mobile platforms.


\ignore{

we believe that the \name methodology is more 
broadly applicable -- even software level detectors that rely on
system calls or SQL query time-series to detect malware can 
benefit from evaluating against {\em intelligent} malware that
obfuscates its payloads' system call traces.

Why architects?
New workload where the app under test actively evades modeling (for a black-list approach)
or actively hides within a parent's model (for a white-list detector).
Lots of interest in research. 
And companies like Intel and Qualcomm 
to build hardware support for malware detection.
Lizy's papers?

WHy not Oakland/CCS? semantic signals v. hardware signals.

We propose \name, a methodology for evaluating mobile malware detectors based
on the principles that malware should correctly execute its payload, that the
payloads should be diversified to pin-point where malware detectors break down,
and that to determine its detection rate and false positives, malware should be compared to
its closest benign app while both apps driven by real user inputs.
In summary, this paper lays a solid methodological foundation for future research into 
novel hardware-based malware detectors.

%



\ignore{

We also analyzed limitations of computational signatures and experimentally
showed that malware can remain undetected but only if it reduces its
intensity and operates less efficiently. Consequently, \name is a tool that 
significantly reduces the rate of damage from widespread infection.

We proposed
and evaluated \name under a realistic scenario approach to construct
effective computational signatures for mobile applications. We believe that our
approach is easily extendable and can be used beyond the realm of mobile apps.
Comparing to the other solutions, the advantage of our method consists of its
whitelist nature.


We clearly showed that malicious workloads exist which are not detectable
using computational signatures. Consequently, computational signatures cannot
replace conventional methods of malware detection used by antivirus companies,
but they can significantly slow down malware dissemination, thus giving
antivirus companies extra time to analyze malware and add its signature into a
database.
}

}

 
{
\bibliographystyle{unsrt}
\bibliography{references}

\begin{thebibliography}{10}

\bibitem{simha-isca}
John Demme, Matthew Maycock, Jared Schmitz, Adrian Tang, Adam Waksman, Simha
  Sethumadhavan, and Salvatore Stolfo.
\newblock On the feasibility of online malware detection with performance
  counters.
\newblock In {\em Proceedings of the 40th Annual International Symposium on
  Computer Architecture}, ISCA '13, pages 559--570, New York, NY, USA, 2013.
  ACM.

\bibitem{anomaly-detector-columbia-raid}
Adrian Tang, Simha Sethumadhavan, and Salvatore~J. Stolfo.
\newblock Unsupervised anomaly-based malware detection using hardware features.
\newblock In {\em Research in Attacks, Intrusions and Defenses - 17th
  International Symposium, {RAID} 2014, Gothenburg, Sweden, September 17-19,
  2014. Proceedings}, pages 109--129, 2014.

\bibitem{map}
Meltem Ozsoy, Caleb Donovick, Iakov Gorelik, Nael Abu-Ghazaleh, and Dmirty
  Ponomarev.
\newblock Malware-aware processors: A framework for efficient online malware
  detection.
\newblock In {\em Proceeding of the 21st International Symposium on High
  Performance Computer Architecture}, 2015.

\bibitem{ponomarev-ensemble-learning}
Khaled Khasawneh, Meltem Ozsoy, Caleb Donovick, Nael Abu-Ghazaleh, and Dmitry
  Ponomarev.
\newblock Ensemble learning for low-level hardware-supported malware detection.
\newblock In {\em 18th International Symposium on Research in Attacks,
  Intrusions and Defenses (RAID)}, 2015.

\bibitem{cids-survey}
Emmanouil Vasilomanolakis, Shankar Karuppayah, Max M\"{u}hlh\"{a}user, and
  Mathias Fischer.
\newblock Taxonomy and survey of collaborative intrusion detection.
\newblock {\em ACM Comput. Surv.}, 47(4):55:1--55:33, May 2015.

\bibitem{cids-Zhou2010}
Chenfeng~Vincent Zhou, Christopher Leckie, and Shanika Karunasekera.
\newblock A survey of coordinated attacks and collaborative intrusion
  detection.
\newblock {\em Computers \& Security}, 29(1):124 -- 140, 2010.

\bibitem{bouncer}
Trendlabs a look at google bouncer.
\newblock http://blog.trendmicro.com/
  trendlabs-security-intelligence/a-look-at-google-bouncer.

\bibitem{sgx-1}
Matthew Hoekstra, Reshma Lal, Pradeep Pappachan, Vinay Phegade, and Juan
  Del~Cuvillo.
\newblock Using innovative instructions to create trustworthy software
  solutions.
\newblock In {\em Proceedings of the 2Nd International Workshop on Hardware and
  Architectural Support for Security and Privacy}, HASP '13, pages 11:1--11:1,
  New York, NY, USA, 2013. ACM.

\bibitem{sgx-2}
Ittai Anati, Shay Gueron, Simon~P Johnson, and Vincent~R Scarlata.
\newblock Innovative technology for cpu based attestation and sealing.
\newblock HASP '13, 2013.

\bibitem{master-key}
Master key vulnerability.
\newblock
  http://blog.trendmicro.com/trendlabs-security-intelligence/trend-micro-solution-for-vulnerability-affecting-nearly-all-android-devices.

\bibitem{taintdroid}
William Enck, Peter Gilbert, Byung-Gon Chun, Landon~P. Cox, Jaeyeon Jung,
  Patrick McDaniel, and Anmol~N. Sheth.
\newblock Taintdroid: An information-flow tracking system for realtime privacy
  monitoring on smartphones.
\newblock In {\em Proceedings of the 9th USENIX Conference on Operating Systems
  Design and Implementation}, OSDI'10, 2010.

\bibitem{flowdroid}
Steven Arzt, Siegfried Rasthofer, Christian Fritz, Eric Bodden, Alexandre
  Bartel, Jacques Klein, Yves Le~Traon, Damien Octeau, and Patrick McDaniel.
\newblock Flowdroid: Precise context, flow, field, object-sensitive and
  lifecycle-aware taint analysis for android apps.
\newblock In {\em Proceedings of the 35th ACM SIGPLAN Conference on Programming
  Language Design and Implementation}, 2014.

\bibitem{dash06}
Denver Dash, Branislav Kveton, John~Mark Agosta, Eve Schooler, Jaideep
  Chandrashekar, Abraham Bachrach, and Alex Newman.
\newblock When gossip is good: Distributed probabilistic inference for
  detection of slow network intrusions.
\newblock In {\em Proceedings of the 21st National Conference on Artificial
  Intelligence - Volume 2}, AAAI'06, pages 1115--1122. AAAI Press, 2006.

\bibitem{dissect_malware_2012}
Y.~Zhou and X.~Jiang.
\newblock Dissecting android malware: Characterization and evolution.
\newblock In {\em Proceeding SP '12 Proceedings of the 2012 IEEE Symposium on
  Security and Privacy}, pages 95--109, 2012.

\bibitem{contagiodump}
Mobile malware database.
\newblock http://contagiominidump.blogspot.com.

\bibitem{obad}
Obad malware.
\newblock
  http://securityintelligence.com/diy-android-malware-analysis-taking-apart-obad-part-1.

\bibitem{geinimi}
Geinimi malware.
\newblock
  https://nakedsecurity.sophos.com/2010/12/31/geinimi-android-trojan-horse-discovered/.

\bibitem{malware.lu}
Malware database.
\newblock http://malware.lu.

\bibitem{virusshare.com}
Malware database.
\newblock http://virusshare.com.

\bibitem{rage-against-the-cage}
Universal android rooting procedure (rage method).
\newblock http://theunlockr.com/
  2010/10/26/universal-android-rooting-procedure-rage-method/.

\bibitem{gingerbreak}
Gingerbreak apk root.
\newblock http://droidmodderx.com/
  gingerbreak-apk-root-your-gingerbread-device.

\bibitem{exploid}
Exploid.
\newblock http://forum.xda-developers.com/showthread. php?t=739874.

\bibitem{ad-vulna}
Vulnerable \& aggressive adware.
\newblock http://www.fireeye.com/
  blog/technical/2013/10/ad-vulna-a-vulnaggressive-vulnerable-aggressive-adware-threatening-millions.html.

\bibitem{webview-attacks}
Erika Chin and David Wagner.
\newblock Bifocals: Analyzing webview vulnerabilities in android applications.
\newblock In {\em Revised Selected Papers of the 14th International Workshop on
  Information Security Applications - Volume 8267}, WISA 2013, pages 138--159,
  New York, NY, USA, 2014. Springer-Verlag New York, Inc.

\bibitem{evernote}
Evernote patches.
\newblock
  http://blog.trendmicro.com/trendlabs-security-intelligence/evernote-patches-vulnerability-in-android-app/.

\bibitem{applocker}
Applock vulnerability.
\newblock
  http://blog.trendmicro.com/trendlabs-security-intelligence/applock-vulnerability-leaves-configuration-files-open-for-exploit.

\bibitem{user-error-permissions}
Adrienne~Porter Felt, Elizabeth Ha, Serge Egelman, Ariel Haney, Erika Chin, and
  David Wagner.
\newblock Android permissions: User attention, comprehension, and behavior.
\newblock Technical Report UCB/EECS-2012-26, EECS Department, University of
  California, Berkeley, Feb 2012.

\bibitem{RAT}
Android rat malware.
\newblock
  http://www.itpro.co.uk/malware/22627/android-rat-malware-invades-mobile-banking-apps.

\bibitem{mobile-bitcoin-miner}
Mobile bitcoin miner.
\newblock {https://blog.lookout.com/blog/2014/04/24/ badlepricon-bitcoin}.

\bibitem{morpheus-hasp}
Mikhail Kazdagli, Ling Huang, Vijay Reddi, and Mohit Tiwari.
\newblock Morpheus: Benchmarking computational diversity in mobile malware.
\newblock In {\em Workshop on Hardware and Architectural Support for Security
  and Privacy}, 2014.

\bibitem{GoogleProGuard}
http://developer.android.com/tools/help/proguard.html.

\bibitem{monkey}
Ui/application exerciser monkey.
\newblock http://developer.android.com/tools/help/monkey.html.

\bibitem{mayur-naik-gatech}
Aravind Machiry, Rohan Tahiliani, and Mayur Naik.
\newblock Dynodroid: An input generation system for android apps.
\newblock In {\em Proceedings of the 2013 9th Joint Meeting on Foundations of
  Software Engineering}, ESEC/FSE 2013, pages 224--234, New York, NY, USA,
  2013. ACM.

\bibitem{reran}
Record and replay for android.
\newblock http://www.androidreran.com.

\bibitem{jitbit}
Jitbit macro recorder.
\newblock http://www.jitbit.com/.

\bibitem{Pelleg-xmeans}
Dan Pelleg and Andrew~W. Moore.
\newblock X-means: Extending k-means with efficient estimation of the number of
  clusters.
\newblock In {\em Proceedings of the 7th International Conference on Machine
  Learning}, 2000.

\bibitem{schapire2012boosting}
R.E. Schapire and Y.~Freund.
\newblock {\em Boosting: Foundations and Algorithms}.
\newblock MIT Press, 2012.

\bibitem{evade-bouncer}
Dissecting android's bouncer.
\newblock https://www.duosecurity.com/ blog/dissecting-androids-bouncer.

\bibitem{chex}
Long Lu, Zhichun Li, Zhenyu Wu, Wenke Lee, and Guofei Jiang.
\newblock Chex: Statically vetting android apps for component hijacking
  vulnerabilities.
\newblock In {\em Proceedings of the 2012 ACM Conference on Computer and
  Communications Security}, CCS '12, pages 229--240, New York, NY, USA, 2012.
  ACM.

\bibitem{mihai-issta}
Davide Canali, Andrea Lanzi, Davide Balzarotti, Christopher Kruegel, Mihai
  Christodorescu, and Engin Kirda.
\newblock A quantitative study of accuracy in system call-based malre
  detection.
\newblock In {\em Proceedings of the 2012 International Symposium on Softre
  Testing and Analysis}, ISSTA 2012, pages 122--132, Neork, NY, USA, 2012. ACM.

\bibitem{android-demystified}
Adrienne~Porter Felt, Erika Chin, Steve Hanna, Dawn Song, and David Wagner.
\newblock Android permissions demystified.
\newblock In {\em Proceedings of the 18th ACM Conference on Computer and
  Communications Security}, CCS '11, pages 627--638, New York, NY, USA, 2011.
  ACM.

\bibitem{DNADroid}
Jonathan Crussell, Clint Gibler, and Hao Chen.
\newblock Attack of the clones: Detecting cloned applications on android
  markets.
\newblock In Sara Foresti, Moti Yung, and Fabio Martinelli, editors, {\em
  Computer Security – ESORICS 2012}, volume 7459 of {\em Lecture Notes in
  Computer Science}, pages 37--54. Springer Berlin Heidelberg, 2012.

\bibitem{DroidMOSS}
Wu~Zhou, Yajin Zhou, Xuxian Jiang, and Peng Ning.
\newblock {Detecting repackaged smartphone applications in third-party android
  marketplaces}.
\newblock In {\em CODASPY '12 Proceedings of the second ACM conference on Data
  and Application Security and Privacy}, pages 317--326, 2012.

\bibitem{Juxtapp}
Steve Hanna, Ling Huang, Edward Wu, Saung Li, Charles Chen, and Dawn Song.
\newblock Juxtapp: A scalable system for detecting code reuse among android
  applications.
\newblock In Ulrich Flegel, Evangelos Markatos, and William Robertson, editors,
  {\em Detection of Intrusions and Malware, and Vulnerability Assessment},
  volume 7591 of {\em Lecture Notes in Computer Science}, pages 62--81.
  Springer Berlin Heidelberg, 2013.

\bibitem{PiggyApp}
Wu~Zhou, Yajin Zhou, Michael Grace, Xuxian Jiang, and Shihong Zou.
\newblock {Fast, scalable detection of "Piggybacked" mobile applications}.
\newblock In {\em CODASPY '13 Proceedings of the third ACM conference on Data
  and application security and privacy}, pages 185--196, 2013.

\bibitem{ernst-taint-type}
Michael~D. Ernst, Ren{\'e} Just, Suzanne Millstein, Werner Dietl, Stuart
  Pernsteiner, Franziska Roesner, Karl Koscher, Paulo~Barros Barros, Ravi
  Bhoraskar, Seungyeop Han, Paul Vines, and Edward~X. Wu.
\newblock Collaborative verification of information flow for a high-assurance
  app store.
\newblock In {\em Proceedings of the 2014 ACM SIGSAC Conference on Computer and
  Communications Security}, CCS '14, pages 1092--1104, New York, NY, USA, 2014.
  ACM.

\bibitem{apposcopy}
Yu~Feng, Saswat Anand, Isil Dillig, and Alex Aiken.
\newblock Apposcopy: Semantics-based detection of android malware through
  static analysis.
\newblock In {\em Proceedings of the 22Nd ACM SIGSOFT International Symposium
  on Foundations of Software Engineering}, FSE 2014, pages 576--587, New York,
  NY, USA, 2014. ACM.

\bibitem{DroidScope}
Lok~Kwong Yan and Heng Yin.
\newblock Droidscope: Seamlessly reconstructing the os and dalvik semantic
  views for dynamic android malware analysis.
\newblock In {\em Proceedings of the 21st USENIX Conference on Security
  Symposium}, Security'12, pages 29--29, Berkeley, CA, USA, 2012. USENIX
  Association.

\bibitem{XuSA2012}
Rubin Xu, Hassen Sa\"{\i}di, and Ross Anderson.
\newblock Aurasium: Practical policy enforcement for android applications.
\newblock In {\em Proceedings of the 21st USENIX Conference on Security
  Symposium}, Security'12, pages 27--27, Berkeley, CA, USA, 2012. USENIX
  Association.

\bibitem{HornyackHJSW2011}
Peter Hornyack, Seungyeop Han, Jaeyeon Jung, Stuart Schechter, and David
  Wetherall.
\newblock These aren't the droids you're looking for: Retrofitting android to
  protect data from imperious applications.
\newblock In {\em Proceedings of the 18th ACM Conference on Computer and
  Communications Security}, CCS '11, pages 639--652, New York, NY, USA, 2011.
  ACM.

\bibitem{paxson1999}
Vern Paxson.
\newblock Bro: A system for detecting network intruders in real-time.
\newblock {\em Comput. Netw.}, 31(23-24):2435--2463, dec 1999.

\end{thebibliography}
}





\end{document}